# Integral Field Spectroscopy of HII region complexes. The outer disk of NGC 6946.

R. García-Benito[1,2]⋆, A. Díaz[2], G. F. Hägele[2,3], E. Pérez-Montero[4], J. López[5], J. M. Vílchez[4], E. Pérez[4], E. Terlevich[5], R. Terlevich[5]†, D. Rosa-González[5]

[1]*Kavli Institute of Astronomy and Astrophysics, Peking University, 100871, Beijing, China*
[2]*Departamento de Física Teórica, Universidad Autónoma de Madrid, 28049 Madrid, Spain*
[3]*Facultad de Cs Astronómicas y Geofísicas, Universidad Nacional de La Plata, Paseo del Bosque s/n, 1900 La Plata, Argentina*
[4]*Instituto de Astrofísica de Andalucía, CSIC, Apartado de correos 3004, 18080 Granada, Spain*
[5]*INAOE, Tonantzintla, Apdo. Postal 51, 72000 Puebla, México*

**ABSTRACT**

Integral Field Spectroscopy obtained with PPak and the 3.5m telescope at the Calar Alto Observatory has been used to study an outer H II region complex in the well studied galaxy NGC 6946. This technique provides detailed maps of the region in different emission lines yielding spatially resolved information about the physical properties of the gas. The configuration was chosen to cover the whole spectrum from 3600 up to 10000 Å, allowing the measurement of the near infrared [S III] lines. We selected four luminous knots, to perform a detailed integrated spectroscopic analysis of these structures and of the whole PPak field-of-view (FOV). For all the knots the electron density has been found to be very similar and below 100 cm$^{-3}$. The [OIII] electron temperature was measured in knots A, B, C and in the integrated PPak-field, and was found to be around 8000 K. The temperatures of [O II] and [S III] were estimated in the four cases. The elemental abundances computed from the "direct method" are typical of high metallicity disk HII regions, with a mean value of 12+log(O/H)= 8.65, comparable to what has been found in this galaxy by other authors for regions at similar galactocentric distance. Therefore, a remarkable abundance uniformity is found despite the different excitations found throughout the nebula. However, due to the quality of the data, the electron temperatures and metallicities obtained, have associated errors comparable to the typical dispersion found in empirical calibrations.

Wolf-Rayet features have been detected in three of the knots, leading to a derived total number of WR stars of 125, 22 and 5, for knots A, C and B, respectively. The ratios of the number of WR to O stars are consistent with the prediction of Starburst99 for individual bursts with an age about 4 Myr. Knot D, with no WR features, shows weak Hα emission, low excitation, and the lowest Hβ equivalent width, all of which points to a more evolved state.

The integrated spectrum of the whole PPak FOV shows high excitation and a relatively evolved age which does not correspond to the individual knot evolutionary stages. Some effects associated to the loss of spatial resolution could also be evidenced by the higher ionising temperature that is deduced from the $\eta$' parameter measured in the integrated PPak spectrum with respect to that of the individual knots.

## 1 INTRODUCTION

NGC 6946 is a relatively nearby, nearly face–on spiral galaxy with an exceptionally gas–rich disk which shows evidence for a high star formation rate (SFR) throughout (Degioia-Eastwood et al. 1984) and has been classified as having strong nuclear starburst activity (Elmegreen et al. 1998). The six historical supernova remnants recorded during the 20th century attest to this high star formation and justify the popular name "Fireworks Galaxy". NGC 6946's orientation and large angular size ($D_{25,B} = 11'.2$) offer the prospect of studying the structure and star forming properties of the disk.

⋆ E-mail: luwen@pku.edu.cn
† Research Affiliate at IoA, University of Cambridge, UK

Morphologically, NGC 6946 is a late-type SAB(rs)cd spiral galaxy (de Vaucouleurs et al. 1991), with several spiral arms and star-forming regions scattered throughout the disk, considerable extinction and a small nucleus. $K$-band images reveal four prominent, not very symmetric arms (Regan & Vogel 1995). NGC 6946 appears in the ARP's atlas, due to the "thick" optical arm in the NE, also seen in the deep H$\alpha$ map presented by Ferguson et al. (1998).

NGC 6946 and its many supernova remnants have been the subject of numerous X–ray, optical and radio studies. At mm wavelengths CO, observations have identified NGC 6946 as having one of the most massive and extended molecular gas components observed in a nearby galaxy (Young et al. 1995). The CO molecule, used as a tracer of the dominant molecular species, $H_2$, has been observed by several authors (Ball et al. 1985; Casoli et al. 1990), showing that the CO distribution has a central density peak and bar-like gas structure. Atomic gas has also been observed throughout an extended disk, although it is found in highest concentrations in the spiral arms and appears less massive than the molecular component (Tacconi & Young 1986). Several studies of the disk H II regions have been made (Hodge 1969; Hodge & Kennicutt 1983). There is a nuclear starburst within the central 11″ (Degioia-Eastwood 1985) with the bright nucleus being characterized as an H II region on the basis of the strength of the [O I]/H$\alpha$ and [S II]/H$\alpha$ line ratios (Ho et al. 1997).

The distance to NGC 6946 has been reported to lie within the range 3.2-11∼Mpc. This uncertainty is partly due to the galaxy being located at a relatively low Galactic latitude (b = 12°) and subject to significant foreground extinction from our Galaxy, quoted from $A_B = 1.48$ (Schlegel et al. 1998) up to $A_B = 1.62$ (Burstein & Heiles 1984). The distance listed in the Nearby Galaxies Catalogue (Tully 1988) is 5.5 Mpc, but a more recent value is $5.9 \pm 0.4$ Mpc (Karachentsev et al. 2000), based on blue supergiants. Here we will use a distance of 5.9 Mpc. Assuming this value, 1″ in the sky corresponds to 28.6 pc.

As mentioned above, NGC 6946 has a particularly extended disk of neutral hydrogen and is among our best examples of galaxies with extreme outer disk star formation. The giant H II regions analysed in this paper are located in the "thick" NE optical arm, at about 4′ (∼ 7 kpc) from the centre of the galaxy. This NE arm is the brightest arm at all wavelengths, indicating a site of vigorous star formation.

Giant Extragalactic H II Regions (GEHRs) observed in many external late-type galaxies, spirals or irregulars, are prominent sites of massive star formation. The large Balmer luminosities of these gigantic volumes of ionized gas attest to the fact that they harbour $10^3$ to $10^5$ $M_\odot$ in OB stars with total ionizing power equivalent to tens or even several hundreds of O5 V stars. The presence of GEHRs in nearby galaxies offer a unique opportunity to study in detail their ionization structure. High spatial resolution imaging has revealed that in many of these objects the ionized material presents a complex structure.

However, the reliability of these findings rests ultimately on the correct interpretation of observations, the main assumption being that the results obtained from the analysis of integrated spectra of a given knot is representative of the whole region. The properties usually derived include: elemental abundances and ages which are subsequently used to characterize the evolutionary state and star formation history of a given galaxy.

There are several reasons to probe the adequacy of the main underlying assumptions that, *i*) the abundances are uniform throughout the whole region and representative of it, and *ii*) there is a unique age of the stellar population of a given region. The derivation of abundances requires the previous knowledge of the physical conditions of the gas, including excitation, density and geometrical effects, and these conditions are known to vary from site to site in a given nebula (*e.g.* Diaz et al. 1987). Integrated spectra are weighted by luminosity and/or surface brightness, therefore the assertion that the abundances of the brightest part of a nebula are representative of the whole needs to be substantiated.

The questions above can only be addressed with the use of spatial information. Until the final decade of the XX century, except for kinematic observations, very little two-dimensional information existed of individual giant extragalactic H II regions that was generally obtained from long slit spectroscopy, mainly for three bright regions in nearby galaxies: NGC 5471 in M101 (Skillman 1985), NGC 604 in M 33 (Diaz et al. 1987), and 30 Doradus in the LMC (Rosa & Mathis 1987). However, with the advent of the HST, high spatial resolution photometry has allowed the study of the high mass stellar content of some regions in nearby galaxies. The results are rather complex showing that, in general, the stellar populations found in giant H II regions include evolved intermediate mass stars (Walborn & Blades 1997). More recently, Integral Field Spectroscopy is revealing itself as a powerful tool to obtain two-dimensional spectrophotometric data which can provide simultaneously information about the physical conditions and abundances of the ionized gas and the stellar content of star forming regions. This technique is also being applied to the study of H II galaxies, but in this case, the spatial resolution applies to the whole galaxy and the resolved units are individual GEHR (*e.g.* Kehrig et al. 2008; Lagos et al. 2009). A two-dimensional analysis of the ionized material in nebulae provides spatially resolved information on properties of the ionized gas. This is important in order to see whether there are local abundance variations within the region and to check if the analysis of gas showing different degrees of ionization yield in fact consistent abundances for locations in different parts of the nebula.

To investigate these issues, we obtained integral field spectroscopy of the large GEHR complex shown in Figure 1. We have produced maps of emission lines, continuum emission and properties of the ionized gas. We have measured the number of Wolf-Rayet stars and traced their location. In the light of all these results, we discuss the issues mentioned before.

In the following section the observations are described. In section 3 the reduction process is described in consider-able detail. In sections 4, 5 and 6 the results are presented. We discuss these results in section 7.

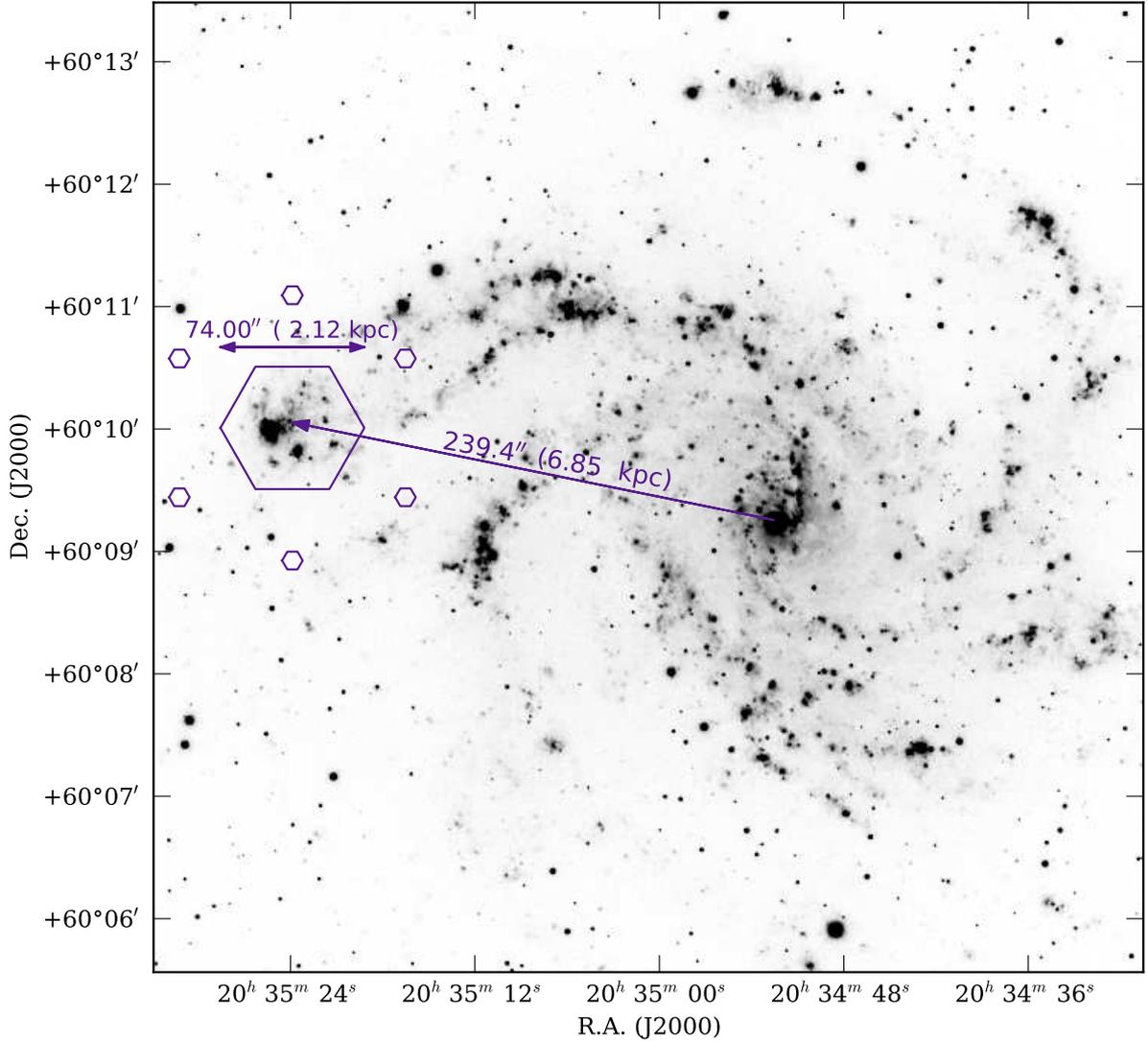

**Figure 1.** Hα image of NGC 6946 taken at the Kitt Peak National Observatory (KPNO) 2.1-meter telescope (Kennicutt et al. 2003). The image has not been continuum-subtracted. The big hexagon shows the position of the PPak field and the little ones, the position of the six bundles of sky-fibers. North is up and East is towards the left-hand side.

## 2 OBSERVATIONS

The group of H II regions of NGC 6946 were observed with the Postdam Multi-Aperture Spectrophotometer (PMAS; Roth et al. 2005), which is attached to the 3.5 m telescope at Calar Alto. The PPak (PMAS fiber Package) fiber bundle IFU (Kelz et al. 2006) was used. PPak spans $74'' \times 64''$ on the sky and provides one of the largest fields of view (FOV) of IFU available worldwide. Therefore, PPak is ideally suited for spectroscopic studies of extended astronomical objects with low surface brightness, such as the outskirts of spiral galaxies.

The PPak IFU holds 331 fibers in a densely packed hexagonal grid with a maximum diameter of $74''$ (which corresponds to approximately 2.1 kpc at the distance of NGC 6946), while each fiber projects to $2''.68$ in diameter on the sky. The fiber-to-fiber pitch is $3''.6$ which yields a total filling factor of ~60%. Additional 36 fibers are distributed in six bundles of 6 fibers each, located following a circular distribution and placed $72''$ from the centre to sample the surrounding sky. Finally, there are 15 extra fibers connected to the calibration unit which are not part of the IFU, and that can be illuminated with light from spectral-line lamps during the science exposures. Therefore, the total number of fibers is 382, distributed in 331 science-fibers, 36 sky-fibers and 15 calibration-fibers.

The data were acquired on 2007 September 7th and 8th during a three nights observing run and under photometric conditions, with a seeing ranging between $0''.8$ and $1''.0$. The airmass of the science images was always less than 1.1 to avoid large Differential Atmospheric Refraction (DAR) effects. The V300 grating was used with two different Grating Rotator angles (GROT). The value GROT=-75 covers a wavelength range between 3700Å and 7000 Å, while the value GROT=-72 spans from 7000 up to 10100 Å. This configuration was chosen in order to cover the whole spectrum from about 3600 to 10000 Å guaranteeing the detection of the [OII] $\lambda\lambda$ 3727,3729 Å and [SIII] $\lambda\lambda$ 9069,9532 Å lines and other emission lines crucial for making plasma diagnostics. The images were taken in the 2×2 pixel binning mode, giving a final dispersion of 3.2 Å/pixel and a resolution of $\sim$10 Å FWHM. Instrumental configuration and details on the exposures taken are given in the journal of observations in Table 1.

**Table 1.** Journal of observations and instrumental configuration. The coordinates of the IFU field centre are $\alpha = 20^h36^m02^s.1$; $\delta = +60°17'12''.9$ (J2000).

| Object | Offset $(\alpha, \delta)$ (arcsec) | Exptime (s) | Grating | GROT | Spec. Range (Å) | Dispersion (Å pix$^{-1}$) | Date |
|---|---|---|---|---|---|---|---|
| NGC 6946 | 0,0 | $3 \times 500$ | V300 | -75 | 3700-7000 | 3.2 | 09/07/2007 |
| NGC 6946 | 1.56,0.78 | $3 \times 500$ | V300 | -75 | 3700-7000 | 3.2 | 09/07/2007 |
| NGC 6946 | 1.56,-0.78 | $3 \times 500$ | V300 | -75 | 3700-7000 | 3.2 | 09/07/2007 |
| NGC 6946 | 0,0 | $3 \times 500$ | V300 | -72 | 7000-10100 | 3.2 | 09/08/2008 |
| NGC 6946 | 1.56,0.78 | $3 \times 500$ | V300 | -72 | 7000-10100 | 3.2 | 09/08/2007 |
| NGC 6946 | 1.56,-0.78 | $3 \times 500$ | V300 | -72 | 7000-10100 | 3.2 | 09/08/2007 |

The coordinates of the IFU field centre were $\alpha = 20^h36^m02^s.1$; $\delta = +60°17'12''.9$ (J2000) and a total of nine single exposures per object and GROT of 500 seconds each were taken. These nine exposures were divided in 3 groups corresponding to different pointings. Each of these three pointings have a small offset in position from the other ones in order to perform a dithering mosaic. This allows to have a final mosaic with a filling factor of 1, so that there is no flux loss. This procedure was also followed in the case of the standard stars. The observed spectrophotometric standard stars were BD +28°4211 for the blue part of the spectrum (GROT=-75) and BD +17°4708 for the red one (GROT=-72). They were used to obtain the characteristic sensitivity function of the telescope and spectrograph for the spectral flux calibration. Calibration images were obtained following the science exposures and consisted of emission line lamps spectra (HgCdHe) or ARC exposures, and spectra of a continuum lamp needed for the wavelength calibration and to locate the spectra on the CCD, respectively.

During the second night, when acquiring the exposures in the 7000-10100 Å range (GROT = -72), the de-rotator and guiding stopped at the second pointing of the dithering exposures, losing the pointing for the second and the third dithering exposures. This prevented us to perform a mosaic covering the whole FOV. We tried to recover the positions of the last dithering, but no convincing results where achieved in the last mosaic, so we ended with only the first dithering for the red part of the spectrum. Nevertheless, this is not important if only emission line ratios instead of absolute fluxes are needed. We will see in section 6, that this problem does not affect our results when dealing with integrated properties. The mosaic in the blue range was built without any problems. The final row-stacked spectra (RSS) blue mosaic contained 993 spectra (3 × 331), while the red RSS frame 331, since it is only the pointing for the first dithering exposure.

Figure 1 shows an H$\alpha$ plus continuum map of NGC 6946 showing the position of the PPak field at the NE arm of the galaxy. The centre of the pointing is located at 4' (6.85 kpc) from the centre of the galaxy.

## 3  DATA REDUCTION

Data reduction was performed using R3D (Sánchez 2006), E3D (Sánchez 2004), PyRAF (a command language for running IRAF[1] tasks based on the Python scripting language), and a self-made Python module (called `pyR3D`) which contains a Python wrapper for R3D, extra tools (as DAR) and other routines.

The steps needed for the PMAS data reduction can be summarized in the following sequential list: 1) pre-reduction; 2) identification of the position of the spectra on the detector along the dispersion axis; 3) extraction of each individual spectrum; 4) distortion correction of the extracted spectra and dispersion correction (wavelength calibration); 5) fiber-to-fiber transmission correction; flux-calibration (including cosmic-ray detection and removal in standard stars); 6) sky-subtraction; 7) mosaic/dither reconstruction; 8) other corrections (DAR).

*Preliminary steps*

The pre-reduction steps consist of all the corrections common to any CCD-based data and can easily be done by pyRAF. This comprises the creation of a master bias (from a combination of several bias frames) and the subtraction of this image from all raw images. The flat-fielding correction is done by the application of the master CCD flat (usually provided by the Calar Alto Observatory) to all the raw images. Next step consists of the combination of different exposures on the same target and same pointing. This also performs the cosmic ray rejection. In the case of the standard stars, we had only one exposure per pointing, so it was necessary to clean for cosmic rays manually. We applied the code L.A. COSMIC written by van Dokkum (2001), using a Laplacian edge detection method. The best set of parameters had been found by eye varying the parameters with respect to the default values until an acceptable result was achieved. Another task which we found to work well was the `lineclean` IRAF routine. We applied both methods to the standard star images and chose the best result for each case.

---

[1] The Image Reduction and Analysis Facility is distributed by the National Optical Astronomy Observatories, which is operated by the Association of Universities for Research in Astronomy, Inc. (AURA) under cooperative agreement with the National Science Foundation (NSF).

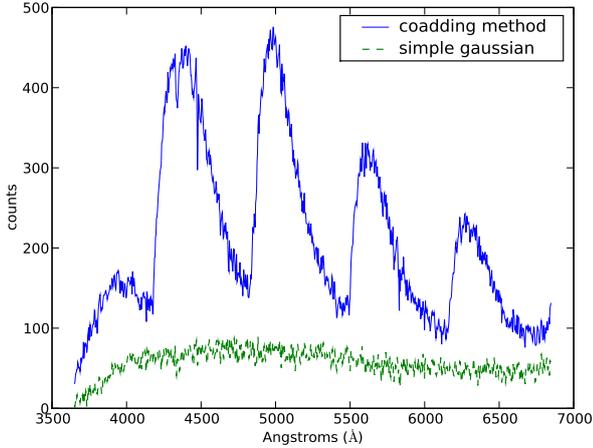

**Figure 2.** Comparison of the coadding (blue continuous line) and simple gaussian (green dot-dashed line) method for the aperture extraction step for a fiber adjacent to a bright source.

*Spectra extraction*

The raw data from fiber-fed spectrographs consist of a collection of spectra, stored as a 2D frame, aligned along the dispersion axis. Each spectrum is also spread along the perpendicular ("cross-dispersion" or "spatial") axis. Spectra are separated by a certain width, following a characteristic profile which may be considered Gaussian. When the spectra are tightly packed, as in PPak mode, contamination occurs among neighbours. This is the so-called cross-talk. It is important to take into account, not only in the raw data but also in the final processed formats, that adjacent spectra at the CCD may originate from distant locations in the sky plane. Special care has to be taken to compensate effects of instrumental flexure, optical distortions, etc, which produce spectra that are not perfectly aligned along the dispersion axis. This effect is corrected by calculating the shifts between the calibration frame and the corresponding object frames. Therefore, it is necessary to find the location of the projection of each spectrum at each wavelength along the CCD in order to extract its corresponding flux. This was done by using continuum illuminated exposures taken at each pointing corresponding to a different orientation of the telescope.

Spectra extraction is usually performed by coadding the flux within a certain aperture around the "trace" of the spectra in the raw data. In PPak data, the aperture extraction normally is done by coadding the flux within an aperture of 5 pixels. Nevertheless, sometimes the aperture extraction is not the optimal method to recover the flux corresponding to each spectrum due to the cross-talk problem and a simple gaussian fitting is preferred. Figure 2 shows a comparison of the coadding and simple gaussian methods for the aperture extraction step in the case of one fiber adjacent to a bright source. The cross-talk problem can affect fibers with low counts adjacent to intense fibers. This is usually seen as a sinusoidal behavior in the continuum of the spectra, producing spurious structure which could lead to wrong results in the interpretation of the data. To avoid the cross-talk problem, sometimes it is enough to reduce the coadding width, but this may produce a substantial loss of flux if the aperture is too small.

In the 2D image resulting from the aperture extraction, the X-axis corresponds to the original dispersion axis, while the Y-axis corresponds to the ordering of the spectra along the pseudo-slit. This is the so-called row-stacked spectra representation (RSS). Each single spectrum corresponds to a particular fiber and the RSS is just a frame, where all the spectra are sorted in the rows of the frame one after another. This representation is very useful because all the information can be stored in a single image, preserving the two-dimensional structure of the data as detected by the CCD plane. However, an extra file is needed (either FITS or ASCII) with additional information of the exact location (projected position) of each fiber in the sky (the "position table").

*Wavelength calibration*

Once the spectra have been extracted and corrected for distortions, we need to wavelength calibrate the observed frames. Wavelength solution is found by identifying the wavelengths of the arc emission lines, using an interactive routine. For the blue spectrum, the arc exposures with a HeHgCd+ThAr lamp had enough strong emission lines to find a good dispersion solution. Unfortunately, the same calibration lamp was used for the red spectra, having only a few intense lines at the lower wavelength end of the range. For this reason, sky lines were used in order to perform a dispersion correction. This is achieved by using the science frames themselves, which contain strong emission sky lines. In the red and near-infrared spectral regions, the OH emission bands of the night-sky spectrum dominate. Osterbrock & Martel (1992) tabulated accurate wavelengths of the individual lines in these bands. Twelve to fifteen lines homogeneously distributed along the full wavelength range were selected, achieving for the calibration an accuracy of 0.5 Å.

*Flux calibration*

After correcting for differences in fiber-to-fiber transmission, we performed the flux calibration of the science frames. A star is a point source on the sky, but the image of the star is broadened due to the turbulence in the atmosphere and/or diffraction of the telescope optics.

The normal procedure for PPak data is to extract the most intense fiber assuming that the star is well centred in a single fiber and that the size of the fiber is bigger than the seeing-disk. However, these assumptions may not be right. Moreover, the PPak mode does not cover the entire FOV, having a filling factor of 60%, which imposes flux losses.

The usual approach to solve the problem is to perform relative spectrophotometry, and recalibrate later using additional information, like broad-band photometry. Alternatively, like for the science frames, one can dither the standard star covering all of the FOV. Once the mosaic is sky-subtracted, spaxels containing the star flux must be selected. It is important to have a smooth final flux ratio curve. Before comparing the 1D count-rate spectrum with the flux table calibration it is advisable to fit a continuum curve to the spectra. This is specially important in the red part (8000-11000 Å range), where absorption features from the atmosphere can depress some parts of the continuum, introducing spurious features, and therefore affecting the final flux ratio curve. It is advisable to have several measurements of different standard stars along the night, or several exposures of the same star at different airmasses in order to have a representative sensitivity function. Then, an average sensitivity function is computed from those with the same instrumental setup. Unfortunately, our data set only had one observation of a standard star per night, so we had to derive a flux ratio for each wavelength range only from one curve. We applied this sensitivity function to the science frames taking into account the airmass and the optical extinction due to the atmosphere in Calar Alto (Sánchez et al. 2007). We also use the "standard" method of flux calibration using only one dithering (both blue and red data) since we could only use the first pointing of the red frames. Despite all these facts, after using both methods for flux calibration, in the overlapping region of the spectra taken for both setups, the agreement in the average continuum level in a fiber-to-fiber comparison was about 5 per cent.

*Sky subtraction*

Many IFUs provide special fibers placed with an offset from the observed object to obtain a clear sky spectrum. When these fibers are not available or the target is surrounded by contaminating emission from adjacent objects, the usual method consists on taking an additional exposure of the nearby sky after or before the exposure of the target itself, taking care that there is no significant contaminating emission from other sources. PPak includes additional fibers that probe the sky far enough from the science FOV to avoid contamination by the object. Using the 36 sky fibers, we created a sky spectra by obtaining the median between a certain number of adjacent spectra, clipping those ones with a flux over a certain threshold of the standard deviation. As seen in Figure 1, the HII complex is located at the extreme of the NE arm of the galaxy. As the sky fibers surround the object, some of them might contain some diffuse nebular emission from the end of the arm contaminating the sky spectra. We compared the flux of the H$\alpha$ line in representative spectra with and without sky subtraction. The difference in the flux was negligible[2]. This reliable, free of nebular emission, sky spectrum was subtracted from the science frames.

*DAR correction*

DAR causes that the blue and red images of an object appear at different positions on the focal plane of the telescope. The amplitude of the shift is a function of zenith distance and of the wavelength separation. The effect is important when fiber-to-fiber intensities of different wavelengths are combined, but negligible if spatially integrated intensities are used. Any DAR correction requires to resample the data spatially, which produces a loss of the initial spatial configuration of the spaxels, changing from one spectrum (wavelength versus flux) per spaxel to a 2D flux image at each wavelength point.

The geometrical size of the spaxels is also relevant. In effect, the smaller the spaxels the more of the flux at different wavelengths end up on separate spaxels. As described before, the fibers on the PPak IFU are circular, of 2$''$.68 diameter, and are separated by 3$''$.6. For PPak data DAR effects become important for airmasses, larger than 1.1 (Sandin et al. 2008).

An empirical DAR correction can be performed by tracing the intensity peak of a reference object in the FOV, like a star or an unresolved point source. To check the effects of DAR, we transformed the blue RSS file to a datacube format, we measured the centroid coordinates of a star in the FOV (IRAF `imcntr` task) and shift each spectral image of the datacube to a common centroid position (IRAF `imshift` task). The position did not change significantly along the whole wavelength range, being the average dispersion $\sim 0.1$ pixels. The fact that all our observations were obtained at airmasses lower than the critical value, combined with the large size of the fibers on the PPak IFU, allows us to neglect DAR corrections. Thus, we can work with RSS files for our set of data.

Finally, we built a single RSS file for the whole mosaic for the blue data, containing 993 spectra, covering from 3650 Å to 7000 Å, and a combined RSS file using the first dithering of blue and red data, with 331 spectra, covering from 3650 Å to 9700 Å.

## 4 DATA ANALYSIS

### 4.1 Subtraction of the underlying population

Underlying stellar populations in star forming regions have several effects in the measurement of the emission lines produced by the ionized gas. Their contribution to the continuum of the galaxy affects the measurement of the equivalent widths of the emission line. On the other hand, Balmer and Paschen emission lines are depressed by the presence of absorption wings of stellar origin (*e.g.* Diaz 1988). All the properties derived from ratios which involved these lines, like ionic abundances or reddening, will be affected.

FIT3D (Sánchez et al. 2006) is a package to fit and deblend emission lines which can handle both RSS images and datacubes. It includes several spectral synthesis routines to fit simple stellar populations (SSPs, or instantaneous bursts) synthetic spectral energy distribution (SED) to the continuum of the object and subtract the underlying population, giving final spectra which, in theory, should contain only the gas emission spectra. Each spectrum from the IFU sample is fitted with synthetic SSP models. As explained in Sánchez et al. (2007), models were created using the GISSEL code (Bruzual & Charlot 2003), assuming a Salpeter IMF, for different ages and metallicities. FIT3D contains 72 models covering a discrete grid of 12 ages (5, 25, 100, 290, 640 Myr, 0.9, 1.4, 2.5, 5, 11, 13 and 17 Gyr), and six metallicities ($Z = 0.0001, 0.0004, 0.004, 0.008, 0.02$ and $0.05$). In order to reduce the number of free parameters, we only used a template of 3 metallicities, namely $Z = 0.02, 0.008$ and $0.004$, with the whole set of ages. The set of metallicities were chosen to have an interval which included the reported metallicity of the region (Ferguson et al. 1998). Each spectrum then was fitted to each of the 36 models. FIT3D re-samples the model to the resolution of the data, convolves it with a certain velocity dispersion, and scales it to match the data set by a $\chi^2$ minimization scheme. In order to fit the continuum, it is necessary to create a mask file providing wavelength intervals of the spectra containing emission lines, Wolf-Rayet bumps, and strong residuals from the sky subtraction and other artifacts. We only performed the fitting on the blue mosaic, *i.e.*, only the spectral region at wavelengths bluer than 7000 Å was used. At redder wavelengths the strong night sky emission lines and the telluric absorptions have strong residuals, so they were masked. In any case, the effect from the underlying population in the emission lines in the red spectral region is negligible.

---

[2] The H$\alpha$ contribution of the sky spectra was less than 10% of the lowest value in the science FOV.

### 4.2 Line measurements

As mentioned above, FIT3D is a package to fit and deblend emission lines. Using this software, we fitted line profiles to each of the 993 blue spectra and 331 red spectra in order to derive the integrated flux of each emission line. The program fits the data with a given model computing a minimization of the reduced $\chi^2$ and using a modified Levenberg-Marquardt algorithm. The data is provided through a configuration file where the model is described. The functions available to build the model are a one dimensional gaussian, a N-order polynomial function and a background spectrum. A single Gaussian was fitted to each emission line, using a low order polynomial function to describe the continuum emission. Instead of fitting the entire wavelength range in a row, we used shorter wavelength ranges for each spectrum that sampled one or a few of the analyzed emission lines. For example, an interval of 4800 to 5100 Å was used for measuring H$\beta$ and [OIII] $\lambda\lambda$ 4959,5007Å. This ensures a characterization of the continuum with the simplest polynomial function, and a way to simplify the fitting procedure. The software also allows the definition of emission line systems. This linking method is useful to fit lines that share some properties (*e.g.*, lines that are kinematically coupled with the same width) or include lines whose line ratio is known (*e.g.*, the ratio between the [OIII] lines $\lambda$ 4959 and $\lambda$ 5007). This was essential for accurate deblending of the lines, when necessary. The statistical errors associated with the observed emission line fluxes have been calculated using the expression:

$$\sigma_l = \sigma_c \sqrt{N\left(1 + \frac{EW}{N\Delta}\right)}$$

where $\sigma_l$ is the error in the observed line flux, $\sigma_c$ represents the standard deviation in a box near the measured emission line and stands for the error in the continuum placement, N is the number of pixels used in the measurement of the line flux, EW is the line equivalent width, and $\Delta$ is the wavelength dispersion in Å per pixel (Gonzalez-Delgado et al. 1994). This expression takes into account the error in the continuum and the photon count statistics of the emission line.

In some cases, the absorption of the atmosphere is so strong that it can depress some emission lines, altering fluxes and ratios needed to obtain the physical properties of the target. This is specially important in the red range of the spectrum. In our case, strong absorptions were present around 9300 Å, rendering [SIII] 9532Å unusable. Thus, we relied on the 9069 Å line when deriving temperatures or using empirical calibrators, using the theoretical ratio 2.44 between both lines.

### 4.3 Map production

Maps are data files stored in 2D fits files, reflecting the original arrangement of the spaxels in the sky. In the case of non regularly gridded IFU or RSS files, it is necessary to interpolate the data, as mentioned in previous sections. We have used the `interpol` routine from E3D with a nearest neighbour interpolation, a grid parameter[3] of $10^{-12}$ and a pixel size of 1″, close to the recommended 1/3 of the original spaxel size. All the maps shown in next sections, unless specifically mentioned, have a North *down* and East left orientation, and a 1″/pixel scale. At the adopted distance of 5.9 Mpc 1″ in the sky corresponds to 28.6 pc.

We have masked all values in the maps with relative errors $\geqslant 25\%$. This is applied to any quantity: flux, ratio or combination of lines used in empirical parameters.

## 5 RESULTS

### 5.1 H$\alpha$ maps and morphology

Figure 3 shows the observed H$\alpha$ flux map created from the RSS blue data. The right panel shows an H$\alpha$ plus continuum image taken at the Kitt Peak National Observatory (KPNO) 2.1-meter telescope (same source as Figure 1) showing the PPak field and with the same orientation as the map of the left panel. From this Figure it is clear that the morphology of the complex of HII regions is well recovered from the IFU data.

Hodge & Kennicutt (1983) (HK) catalogued more than 10000 HII regions in 125 galaxies. This HK catalog contains 540 HII regions identified in NGC 6946. We have labelled the knots following the HK identification. As it can be seen in the H$\alpha$ map, only the most intense knots are clearly defined, and for this reason we have labelled them with the first four alphabetical letters, being HK83-003 knot A, HK83-004 knot C and HK83-016 knot D. Interestingly, knot B is not identified in Hodge & Kennicutt (1983). Since they have enough signal-to-noise, we will present the main results for these four knots when discussing the integrated properties. It has to be remarked that the right panel image has not been continuum-subtracted, so some knots are seen more intense since their continuum sources are included in the image, but their H$\alpha$ emission is not so strong. This is the case of knots HK83-013, HK83-014, HK83-009 and HK83-006.

Figure 4 shows the continuum flux around 4250 Å (*left*), near H$\alpha$ (*middle*) and the continuum emission map 8525 Å (*right*) (50 Å width) together with the H$\alpha$ isocontours overplotted as reference. The red map (and its isocontour) comes from the first pointing. Maps of the continua are representative of the stellar emission, free of contamination from the gaseous emission lines. From the Hα isocontours it can be seen that the brightest knot, A, has the most extended structure, with a tail pointing to the north-west. This structure seems to be formed following the continuum morphology. From the continuum emission, it can be seen that the structure of knot A coincides with that of the Hα map, with a small concentration to the west of the knot, probably it being the source of the north-west tail of this knot.

---

[3] The grid parameter represents the resolution of the triangulation.

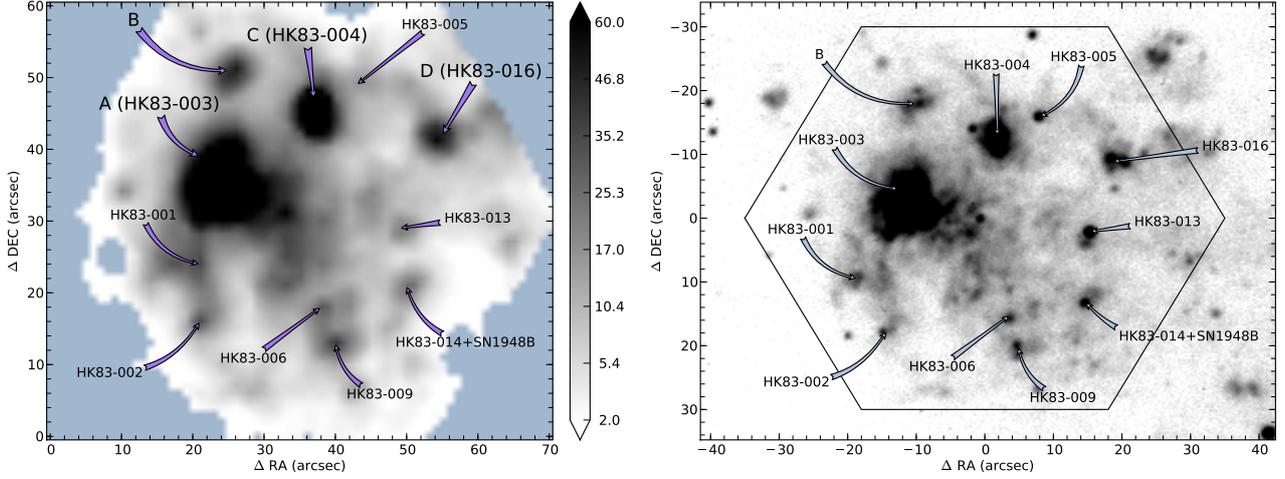

**Figure 3.** *Left panel:* Hα flux map (not corrected for reddening) obtained from the IFU data. The scale bar (valid only for the left panel) is given in units of $10^{-16}$ erg s$^{-1}$ cm$^{-2}$. North is *down* and east is left. The masked values follow the criteria explained in the text. *Right panel:* Hα image taken at the Kitt Peak National Observatory (KPNO) 2.1-meter telescope (same source as Figure 1). The image has not been continuum-subtracted. The hexagon shows the position of the PPak FOV. The orientation is the same as for the left panel. The most important knots are labelled following the identification by Hodge & Kennicutt (1983). The four main intense knots are also labelled with the first alphabet letters. Knot B is not identified in Hodge & Kennicutt (1983).

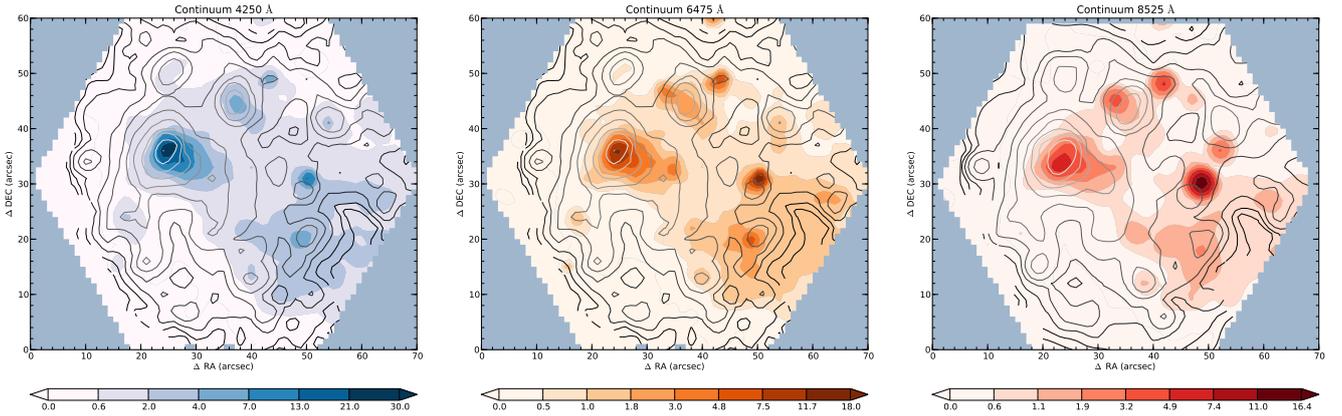

**Figure 4.** Maps in the continuum around 4250 Å (*left*), near Hα (*middle*) and in the continuum near 8525 Å (*right*) (50 Å width) together with the Hα isocontours overplotted. The last map and its isocountours come from the first pointing. Contours are masked according to their emission line flux. The scale bar is given in units of $10^{-17}$ erg s$^{-1}$ cm$^{-2}$.

Although the other knots have their equivalence between their emission line and continuum maps, their maximum position are not the same. A remarkable structure is that of knot C, which has an elongated continuum in a south-east to north-west direction, while the emission line morphology is ellipsoidal. A continuum structure near this knot has the associated identification HK83-005, but it is weak in our emission line maps.

The opposite happens to knot B. Although in the emission line maps some structure is seen, in the continuum maps only a weak blue contribution is present. Knot D has a clear correspondence between the blue continuum and the Hα emission.

Interestingly, knots HK83-013 and HK83-014 have strong continuum contribution, but very low Hα emission, and their emission peaks have an offset as compared to their respective continua. This effect may be due to the fact that the underlying population of these knots is slightly older. Note that knot HK83-014 has an associated supernova (Mayall 1948).

As for the continuum emission near 8500 Å the morphology is very similar to that of the blue part. In fact, the continuum peak of the most important knots in the red matches those from the blue. In some structures, the intensity of the red continuum is more marked, as the case of the source between knots HK83-013 and HK83-016. Knot HK83-013 shows a prominent absorption Ca II triplet ($\lambda\lambda 8498, 8542, 8662$ Å) feature in the red part, probably due to the presence of some luminous red supergiants. The other knot which also shows this feature is HK83-004, although its intensity is very weak.

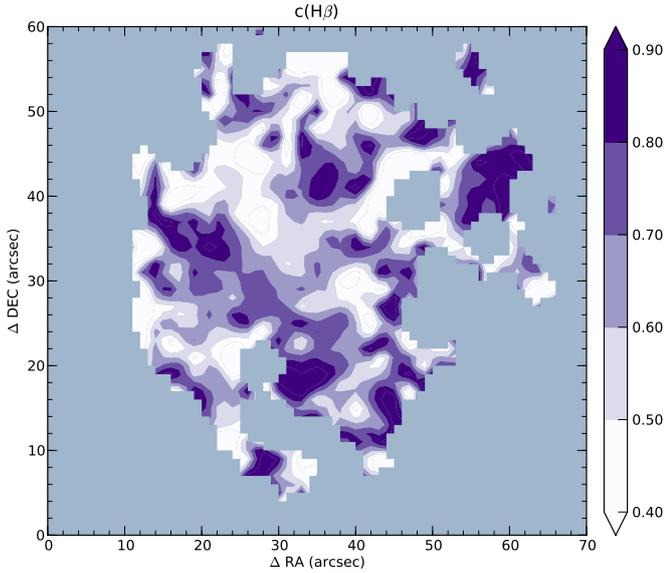

**Figure 5.** c(Hβ) map derived from Hα/Hβ under case B recombination, and assuming the Cardelli et al. (1989) extinction law.

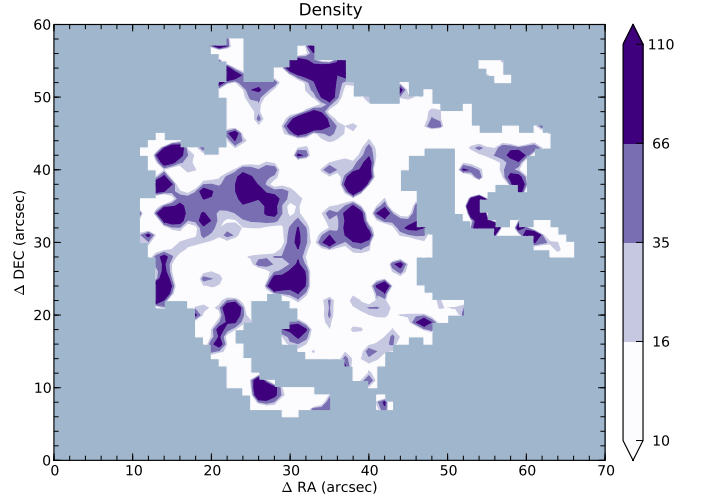

**Figure 6.** Map of the electron density, $n_e$, calculated from the ratio of [S II] $\lambda\lambda$ 6717,6731 Å lines. The scale is in units of cm$^{-3}$.

Apart from the above differences, the overall continuum morphology of the whole region matches the one of the emission line maps, where the complex extends through the west region of the FOV, while the surroundings of the east (south and north) region have lower surface brightness. This is clear in Figure 1, where it is seen that the region is located at the end of the NE arm of NGC 6946.

### 5.2 Reddening correction and c(Hβ) map

Due to low signal-to-noise in Hγ and Hδ, we used only the value of the Balmer decrement derived from Hα/Hβ to estimate the reddening coefficient, c(Hβ) for each fiber spectrum. For the same reason, the most important Paschen lines could not be measured with an acceptable level of precision. For the integrated spectra (see below), we used an iterative method to estimate density, temperature and reddening, taking as starting values those derived from the measured [S II] $\lambda\lambda$ 6717,6731 Å and [O III] $\lambda\lambda$ 4363,4959,5007 Å. The theoretical values have been calculated based on the data by Storey & Hummer (1995). We use a mean value of $n_e = 10^2$ $cm^{-3}$ and $T_e = 8000$ K as characteristic values for the region to obtain the reddening map and the extinction law given by Cardelli et al. (1989) with R$_V$ = 3.1.

The resulting reddening map is shown in Figure 5. The mean and standard deviation for c(Hβ) over the FOV is 0.61 ± 0.19; of this, c(Hβ) ∼ 0.4 (one visual magnitude) comes from Galactic extinction, because NGC 6946 is at a low Galactic latitude. Our mean value is remarkably close to the value found by Ferguson et al. (1998) for knot D (HK16 or FGW 6946B in their nomenclature) of 0.68 ± 0.11. The reddening distribution is consistent with the distribution of the main four knots, where these knots exhibit high values of extinction. Interestingly, high values are also found in the north-west side of knot A, not correlating with any important feature in continuum or emission line maps. It is also remarkable a "stream" of low reddening values which goes east-west between knot A and the other three knots.

For each fiber spectrum we derived its corresponding reddening coefficient and all fluxes of the emission lines (for each fiber) were corrected for extinction using their corresponding c(Hβ) value, which is an important point when deriving the ionization structure and physical-chemical parameters.

### 5.3 Electron density

The electron density of the ionized gas, and other physical conditions (such as temperatures when available), have been derived from the emission line data using the same procedures as in Pérez-Montero & Díaz (2003), based on the five-level statistical equilibrium atom approximation in the IRAF task `temden` (De Robertis et al. 1987; Shaw & Dufour 1995). See Hägele et al. (2008) for a description of the relations of the physical conditions of the gas and ionic abundances. We have taken as sources of error the uncertainties associated with the measurement of the emission-line fluxes and the reddening correction, and we have propagated them through our calculations.

Figure 6 shows a plot of the spatial variation of electron density, $n_e$, as derived from the ratio of the [S II] $\lambda\lambda$ 6717,6731 Å lines (representative of the low-excitation zone of the ionized gas), assuming a temperature of 8000 K. The map shows an overall low density throughout the region, with most of the values below 100 cm$^{-3}$, and presenting a few knots denser than average. One of them shows the extended region of knot A, while two denser zones surround knot C. At any rate, the errors involved in some of the other regions are compatible with a general constant density, well below the critical value for collisional de-excitation. This low nominal value is consistent with other values from the literature (Ferguson et al. 1998). Values lower than ∼ 30 in Figure 6 are numerical results, representative of the low density limit, and are shown just for comparison purposes.

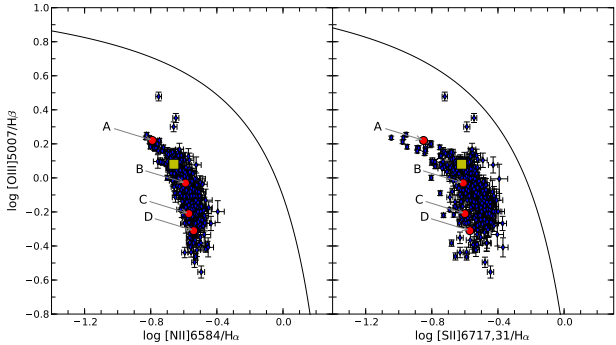

**Figure 7.** Diagnostic diagrams derived from different line ratios, including the values of the integrated spectra (see Section 6). The (red) circle points represent the values of the knots, while the (yellow) square points the position in the diagram of the whole PPak field. *Left panel:* Relation between [O III] $\lambda$ 5007/H$\beta$ and [N II] $\lambda$ 6584/H$\alpha$. *Right panel:* Relation between [O III] $\lambda$ 5007/H$\beta$ and [S II] $\lambda\lambda$ 6717,6731/H$\alpha$. The solid curve represents the theoretical maximum starburst line (Kewley et al. 2001).

In terms of the [S II] $\lambda\lambda$ 6717/6731 ratio, the field average ratio is 1.41 $\pm$ 0.12 (1$\sigma$ error).

### 5.4 Ionization structure and excitation

The hardness (fraction of high energy photons) of the radiation and the density of the gas are two of the main factors that drive the different ionization states of the metals. A "hard" spectrum increases the ionization degree, and this hardness increases with the temperature of the star.

The ionization degree should be measured estimating the ionic abundance of an element. In turn, this implies the estimation of temperature, and the measurement of very low intensity lines. The calculation of the ionic abundance at each point to study the ionization degree throughout the nebula is not feasible. An alternative is to measure the so-called diagnostic ratios. These diagnostics involve two or three strong emission lines whose ratios depend strongly on the ionization degree and, to a lesser extent, on temperature or abundance. The most common ones are [O III] $\lambda$ 5007/H$\beta$, [N II] $\lambda$ 6584/H$\alpha$ and [S II] $\lambda\lambda$ 6717,6731/H$\alpha$. The first ratio is related to highly ionized gas and is a good indicator of the mean level of ionization (radiation field strength) and temperature of the gas, while the last two ratios are related to low ionized gas and are indicators of the number of ionizations per unit volume (ionization parameter).

Diagnostic diagrams (BPT, Baldwin et al. 1981; Veilleux & Osterbrock 1987) can be used to distinguish among possible ionization sources, AGN, shocks or massive stars in star-forming regions.

Figure 7 shows [O III] $\lambda$ 5007/H$\beta$ versus [N II] $\lambda$ 6584/H$\alpha$ and [O III] $\lambda$ 5007/H$\beta$ versus [S II] $\lambda\lambda$ 6717,6731/H$\alpha$ nebular diagnostic diagrams for all measurements in all IFU positions, including the values of the integrated spectra (see Section 6). On these plots, we also show the theoretical "maximum starburst line" derived by Kewley et al. (2001), indicating a conservative theoretical limit for pure photoionization from the hardest starburst ionizing spectrum that can be produced. For points to exist above or to the right-hand side of this threshold, an additional contribution to the excitation from non-stellar sources is required. From these classical BPT diagnostic diagrams it is clear that the line ratios for all the position in the region are located in the general locus of H II-region like objects.

It should be noted the variation of the [O III] $\lambda$ 5007/H$\beta$ ratio across the region, of about 0.8 dex. This is comparable to the spread found in H II galaxies, as in the case of IIZw70 (Kehrig et al. 2008). On the other hand, the variation in the [N II] $\lambda$ 6584/H$\alpha$ and [S II] $\lambda\lambda$ 6717,6731/H$\alpha$ ratios are lower. Numerically, the field average ratio of log([O III] $\lambda$ 5007/H$\beta$) = -0.08 $\pm$ 0.16, log([N II] $\lambda$ 6584/H$\alpha$) = -0.58 $\pm$ 0.07 and log([S II] $\lambda\lambda$ 6717,6731/H$\alpha$) = -0.54 $\pm$ 0.10 (1$\sigma$ errors).

A spatial representation of these three line ratios is shown in Figure 8. For the sake of clarity and comparison, H$\alpha$ is also included in the upper left panel, corrected for reddening. Unlike the reddening map, the structures that appear in the excitation maps are much simpler. In terms of the main four knots, particularly knot A, there is a very good correlation between the value of the intensity of H$\alpha$ and the value of every ratio, the sign of the correlation being negative for log([S II] $\lambda\lambda$ 6717,6731/H$\alpha$) and log([N II] $\lambda$ 6584/H$\alpha$) ratios and positive for the log([O III] $\lambda$ 5007/H$\beta$) ratio. This is the expected behaviour if in the central zones of the H II region sulphur, nitrogen, and oxygen were twice ionized, while only singly ionized in the surroundings due to the increasing distance from the ionizing source.

### 5.5 Wolf-Rayet stellar population

Spectral features specifically produced by WR stars are frequently detected in regions of intense star formation. These WR features are the broad blue bump, centred at 4650 Å and produced mainly by broad emission lines of N V at 4605, 4620 Å, N III 4634, 4640 Å, C III/IV 4650, 4658 Å and He II at 4686 Å, and the red bump, usually fainter, is centred at 5808 Å produced mainly by C III/IV.

We have detected the blue WR bump in several fibers on the IFU data, while the red bump was only marginally detected in the integrated emission of the knot A (Figure 9). The blue bump is remarkably strong in the same integrated spectrum. The integrated properties of the knots are described in section 6.

Figure 10 shows the spatial distribution in the PPak field of the flux associated with the presence of WR stars. The filled dark regions represent the fiber detections of the blue WR bump. On the left panel the H$\alpha$ contours are overplotted, while in the right panel the continuum contours near the WR emission ($\sim \lambda$ 4500 Å) are shown. The filled regions only represent the location of the WR stars in the field. Clear detections were found in 10 fibers of knot A, corresponding to the extended region in that location. On the other hand, a weak blue bump was found only in one fiber of knot B, while in knot C two fibers showed a moderate intensity feature. As it is clearly seen, the location of the bumps follows closely the morphology of the continuum emission adjacent to the WR blue bump. We also verified that the maximum of the continuum emission adjacent to H$\beta$ corresponds spatially to the WR bump intensity maxima. Interestingly, knot B shows a weak WR blue bump. Although the continuum there has very low surface brightness, the local maximum matches the position of the WR.

The flux in the WR bumps have been measured in the following way. First a continuum shape has been adopted and we have measured the broad emission over this continuum in the wavelengths of the blue bump, at 4650 Å. Finally, we have subtracted the emission of the narrower lines not emitted by the WR winds, namely [FeIII] 4658 Å, HeII 4686 Å, HeI + [ArIV] 4711 Å and [ArIV] 4740 Å in the blue bump (although these last two lines present very small contribution). The dereddened relative intensities and the equivalent widths of the blue bump are shown in Table 2. Although some contribution of the red bump was found in knots A and C, the flux could not be measured with an acceptable level of precision. Knot D shows no presence of either blue or red bump. All measurements were performed in the integrated spectra from the mosaic, so that the reported values have been derived from absolute fluxes.

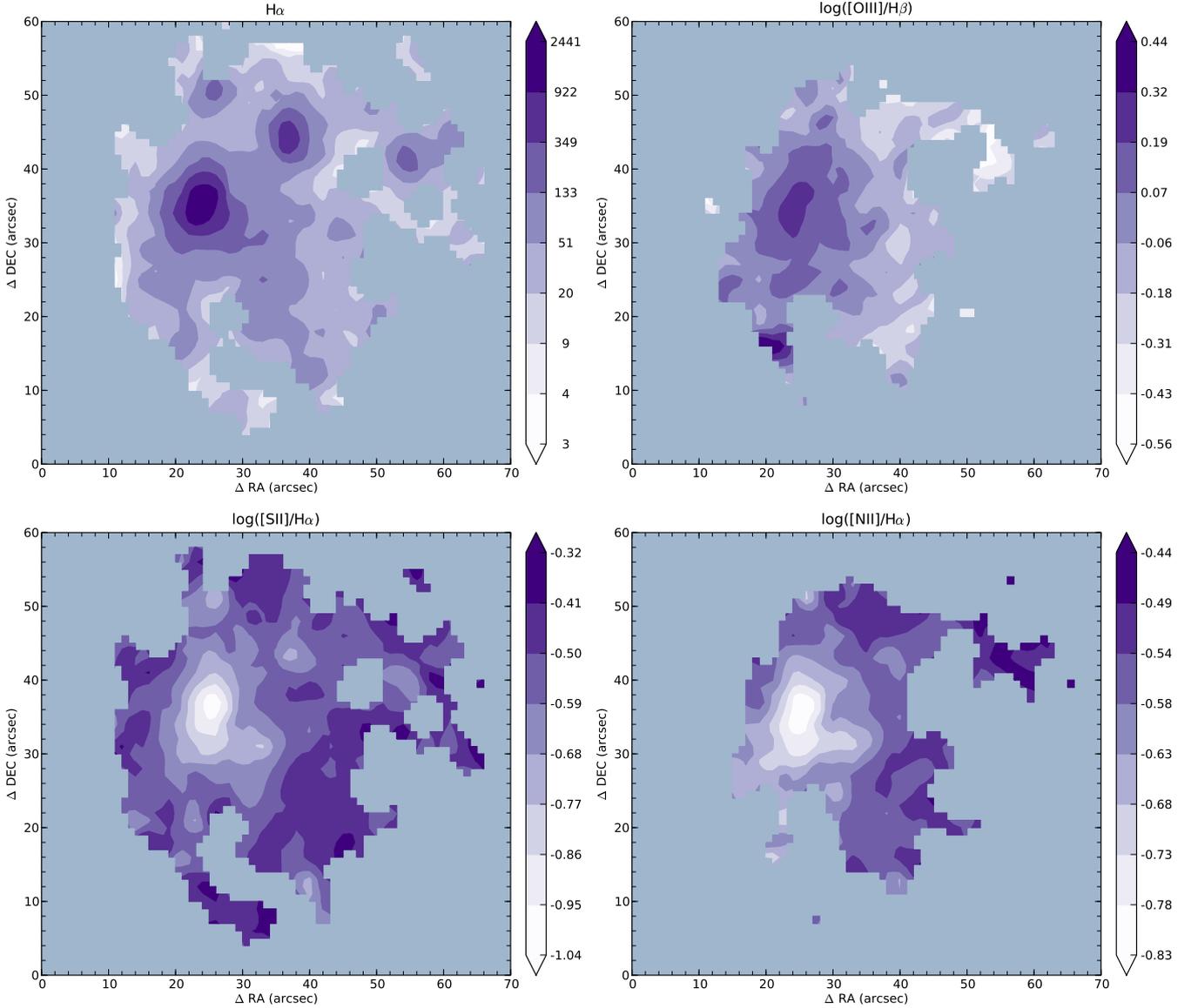

**Figure 8.** *Upper left panel:* Hα flux map (corrected for reddening). The scale bar is given in units of $10^{-16}$ erg s$^{-1}$ cm$^{-2}$. This map is included as reference for the excitation maps. *Upper right panel:* Flux ratio map of log([OIII] λ 5007/Hβ). *Lower left panel:* Flux ratio map of log([SII] λλ 6717,6731/Hα). *Lower right panel:* Flux ratio map of log([NII] λ 6584/Hα).

There are several uncertainties in the derivation of the WR properties, like the effects of underlying stellar population, the nebular emission to the continuum and the dust absorption of the UV (see Pérez-Montero et al. 2010). Thus, all reported WR equivalent widths represent lower limits while relative intensities are upper limits.

## 6 ANALYSIS OF INTEGRATED SPECTRA

In order to study the mean properties of the HII region complex, several integrated spectra, corresponding to the more intense H$\alpha$ emitting knots, were obtained from the PPak-field. These knots, identified in Figure 3, have enough signal-to-noise to allow an analysis of the physical properties of the gas and their abundances by standard methods. Unfortunately, auroral line detections with acceptable signal were not found in individual fibers; the necessary signal-to-noise to measure these weak lines was achieved only in the integrated spectrum of each knot.

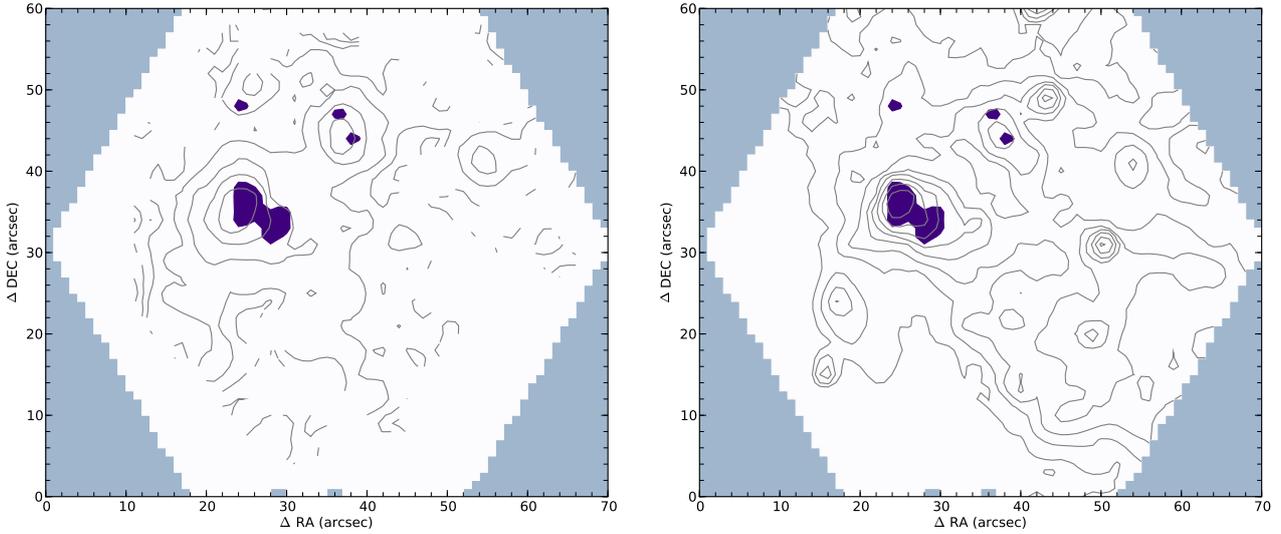

**Figure 10.** Wolf-Rayet (WR) spatial distribution. The filled dark (violet) regions represent the fiber detections of the blue WR bump. On the left panel the H$\alpha$ contours are overplotted, while in the right panel the continuum contours near the WR emission ($\sim \lambda$ 4500 Å) are shown. The filled regions only represent the positions of the WR stars in the field, not their intensity.

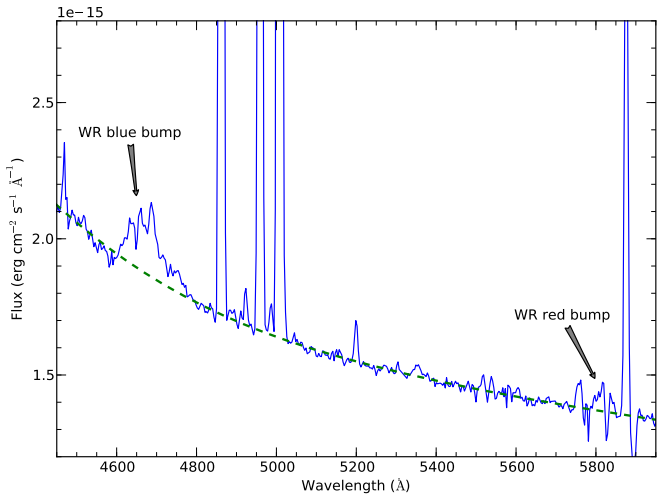

**Figure 9.** Detail of the integrated spectrum of knot A with the identification of both the blue (4650Å) and red (5808Å) Wolf-Rayet bumps over the adopted continuum, marked as a dashed (green) line.

**Table 2.** Luminosities, relative intensities and equivalent widths of the Wolf-Rayet features measured on the integrated spectra of four knots at 4650 Å (blue bump, bb). All quantities have been corrected for reddening.

| ID | log L(bb) (erg · s$^{-1}$) | I(bb)$^a$ | EW(bb) Å |
|---|---|---|---|
| Knot A | 38.58 ± 0.04 | 20.3 ± 1.7 | 14.0 ± 1.2 |
| Knot B | 37.41 ± 0.06 | 27.4 ± 3.5 | 25.5 ± 3.3 |
| Knot C | 37.90 ± 0.05 | 22.7 ± 2.8 | 16.4 ± 2.9 |

$^a$ In units of 100 · I(H$\beta$)

As detailed in the previous section, a complete mosaic of 993 spectra was accomplished for the blue spectral range (3700-7000 Å), while for the red part of the spectrum (7000-10000 Å) we could only use one pointing of 331 spectra. To take advantage of all the relevant emission lines available, we had to build the integrated spectrum for each knot from the first blue and red pointing. Since the FOV for one pointing is not completely covered by all the fibers, we cannot expect to obtain absolute fluxes. Nevertheless, we can assume that the relative line intensities of the lines are maintained, and therefore the derived properties and abundances should be reliable.

The spatial limits of each of the four knots were defined in the H$\alpha$ map by means of visual inspection with the help of iso-contour plots. The outer limit contour is at 20-30% of the maximum peak. To improve the contour definition, we used the H$\alpha$ map built from the mosaic (3 pointings). Once the knots were defined, we co-added all the fibers belonging to each knot in the mosaic exposure, obtaining four absolute flux calibrated integrated spectra. Since each fiber is labelled in both the mosaic and each pointing data, we identified the fibers coming from the first blue and red pointings and co-added them for each knot. Finally, the integrated spectrum over the full FOV of our IFS dataset was built both from the mosaic and from the one pointing data. Thus, we ended up with ten spectra: five absolute flux calibrated spectra from the mosaic data covering the range 3700-7000 Å for the four main knots and the whole FOV, and other five spectra of the same areas coming from the blue and red first pointing covering the range 3700-10000 Å. All the fibers containing WR features were included in these integrated spectra, since the WR positions are near the peak of the Hα emission, as shown in Figure 10.

The spectra of the four knots (labelled from A to D) and the integrated spectrum of the whole PPak-field with some of the relevant identified emission lines are shown in Figure 11. The spectrum of each knot is split into two panels. All of them have been obtained from the first blue and red pointings.

Line fluxes for the most relevant emission lines were measured in each spectrum using the `splot` task in IRAF following the procedure described in Hägele et al. (2006). A pseudo-continuum has been defined at the base of the hydrogen emission lines to measure the line intensities. This procedure, for our particular case, gives the same results, within the observational errors, as fitting synthetic SSP models to the continuum. Since we had to measure the auroral lines manually, we decided to follow the same procedure also for the rest of the lines. The statistical errors associated with the observed emission fluxes have been calculated using the same procedure as described before.

The reddening coefficient c(Hβ) was calculated using the procedures described in Section 5.2. To follow a similar approach as when deriving the maps, we estimated c(Hβ) only by means of Hα and Hβ.

Table 3 gives the equivalent widths and the reddening-corrected emission line fluxes for six integrated spectra, two corresponding to the whole FOV and one corresponding to each of the four emitting knots, together with the reddening constant and its error, and the reddening-corrected Hβ intensity. The adopted reddening curve, $f(\lambda)$, normalized to Hβ, is given in column 2 of each table. The errors in the emission line ratios were obtained by propagating in quadrature the observational errors in the emission-line fluxes and the reddening constant uncertainties. As can be seen, the relative intensities and equivalent widths for the integrated spectra from the entire PPak FOV for the mosaic exposure (PPak-field Mosaic) and for the blue and red first pointing (PPak-field Pointing-1) are in very good agreement, within the errors. This is also the case for the blue spectra of each of the four knots, hence we only show the measured line intensities coming from the first pointing exposure.

In the four knots the total reddening-corrected Hβ intensity ($I_{tot}$(Hβ)) measured in the mosaic is also given. As it can be appreciated, the total flux is between a factor of 2 to 3 higher than the value measured in the spectra corresponding to the single pointing. The Hβ flux of the emission knots amounts to about 50% of the total PPak field-of-view flux, with knot A, B, C, and D contributing 38%, 2%, 7%, and 3% to the total flux, respectively. Knot A is the brightest of them and provides 75% of the combined flux from the four knots.

### 6.1 Electron densities and temperatures

The physical conditions of the ionized gas, including electron temperature and density, were computed using the five-level statistical equilibrium atom approximation in the task `temden` of the IRAF/STSDAS package. For details of the equations involved, the reader is referred to Hägele et al. (2008). The quoted errors are the result of propagating through our calculations the uncertainties associated with the measurement of the emission-line fluxes and reddening correction.

Electron densities are derived from the [SII] λλ 6717,6731 Å line ratio, which is representative of the low-excitation zone of the ionized gas. In all cases, the density is found to be lower than 100 cm$^{-3}$, well below the critical density for collisional de-excitation. As compared to the map of Figure 6, the densities derived from the integrated spectra match well the ones obtained from the map for each particular region.

We computed three electron temperatures: T([OII]), T([OIII]) and T([SIII]) for each of the three knots A, B, and C. For knot D no temperatures were calculated as the auroral line was not detected. The spectral resolution is at the limit to resolve the [OII] λλ 7319,7330 Å doublet lines, so we only measured the integrated flux of the doublet. These lines can have a contribution by direct recombination which increases with temperature (Liu et al. 2000). Using the calculated [OIII] electron temperatures, we have estimated this contribution to be less than 3 per cent in all cases and therefore we have not corrected for this effect. For knot B, the intensity of the [OII] λλ 7319,7330 Å lines did not allow an accurate measure, therefore we derived its [OII] temperature from T([OIII]) using the relation based on the photoionization models described in Pérez-Montero & Díaz (2003), which take into account explicitly the dependence of T([OII]) on the electron density, n:

$$t([\text{OII}]) = \frac{1.2 + 0.002 \cdot n + \frac{4.2}{n}}{t([\text{OIII}])^{-1} + 0.08 + 0.003 \cdot n + \frac{2.5}{n}}$$

Although the [OIII] λ 4363 auroral Å line was detected in the one dithering exposure, the signal-to-noise of the mosaic was higher and the estimated errors smaller. For this reason, we have used the T([OIII]) derived from the mosaic in the abundance analysis. At any rate, the [OIII] temperatures calculated from both sets of data coincide within the errors.

The [SIII] λ 6312 Å auroral line could be measured with enough precision in the integrated spectrum of knot A. The line could not be measured in the remaining knots, so for them the [SIII] temperature was estimated from the empirical relation:

$$t([\text{SIII}]) = (1.19 \pm 0.08)t([\text{OIII}]) - (0.32 \pm 0.10)$$

found by Hägele et al. (2006).

The electron densities and temperatures derived from the integrated spectra of the PPak-field and the four knots are listed in Table 4 along with their corresponding errors.

### 6.2 Chemical abundances

We have derived the ionic abundances of the different chemical species using the stronger emission lines available detected in the spectra and the task `ionic` of the STSDAS package in IRAF (see Hägele et al. 2008). The total abundances have been derived by taking into account, when required, the unseen ionization stages of each element, resorting to the most widely accepted ICFs for each species:

$$\frac{X}{H} = ICF(X^i)\frac{X^{+i}}{H^+}$$

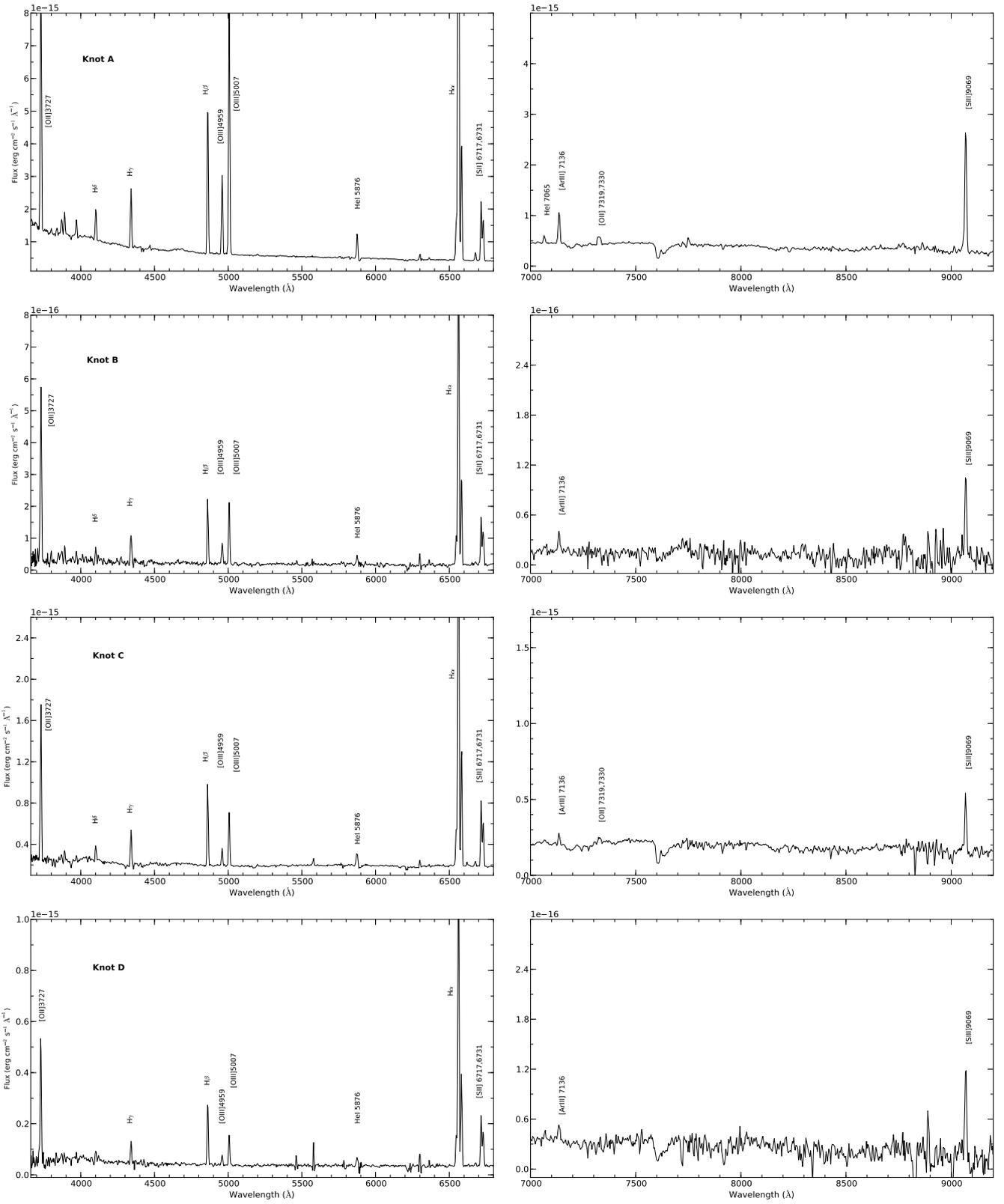

**Figure 11.** Blue and red spectra for the knots A, B, C, D and the integrated PPak-field.

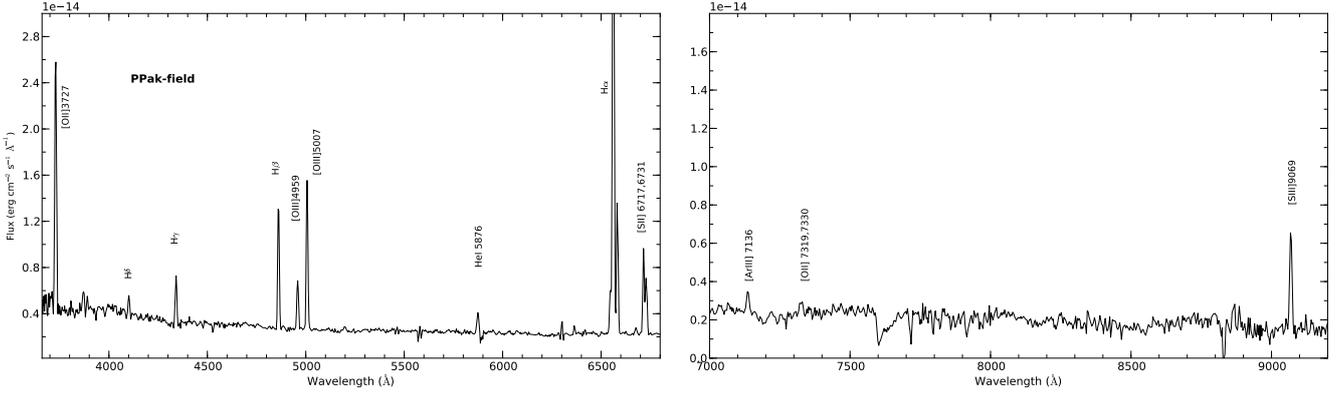

**Figure 11.** (*continued*) Blue and red spectra for the knots A, B, C, D and the integrated PPak-field.

### 6.2.1 Helium abundance

We were able to measure the fluxes of the 4 strongest helium emission lines, HeI $\lambda\lambda$ 4471 (in knot A), 5876, 6678 and 7065 of HeI and HeII $\lambda$ 4686, the latter probably associated with the presence of WR stars. The spectra presented other HeI lines, although all of them too weak to be used to derive abundances with the necessary accuracy.

Helium lines arise mainly from pure recombination; however, they could have some contribution from collisional excitation as well as be affected by self-absorption and, if present, by underlying stellar absorption (see Olive & Skillman 2001, for a complete treatment). We have taken the electron temperature of [OIII] as representative of the zone where the He emission arises and we have used the equations given by Olive & Skillman (2001) to derive the $He^+/H^+$ value. We have not taken into account, however, the underlying stellar population nor the corrections for fluorescence, since the involved helium lines have a negligible dependence on optical depth effects and the observed objects have low densities.

The results obtained for each line and their corresponding errors are presented in Table 5, along with the adopted value for $He^+/H^+$ (a weighted average of the values deduced from each of the lines, using the error of each line as weigth).

### 6.2.2 Ionic and total elemental abundances from forbidden lines

The oxygen ionic abundance ratios, $O^+/H^+$ and $O^{2+}/H^+$, were derived from the [OII] $\lambda\lambda$ 3727,3729 Å and [OIII] $\lambda\lambda$ 4959, 5007 Å lines, respectively using the appropriate electron temperature for each ion.

We derived $S^+$ abundances from the fluxes of the [SII] emission lines at $\lambda\lambda$ 6717, 6731 Å assuming that T([SII]) $\approx$ T([OII]); $S^{2+}$ abundances have been derived from the fluxes of the near-IR [SIII] $\lambda$ 9069 line and the directly measured T([SIII]) for knot A and its estimated value for the other knots. The total sulphur abundance was calculated using an ionization correction factor (ICF) for $S^+$+$S^{2+}$ according to Barker (1980) formula (see Hägele et al. 2008).

The ionic abundance of nitrogen, $N^+/H^+$, was derived from the intensities of the $\lambda\lambda$ 6548, 6584 Å lines assuming that T([NII]) $\approx$ T([OII]).

Neon is only visible in the spectra via the [NeIII] emission line at $\lambda$3869 Å. For this ion, we took the electron temperature of [OIII], as representative of the high excitation zone. The used ICF can be found in Pérez-Montero et al. (2007).

The only accessible emission lines of argon in the optical spectra of ionized regions correspond to $Ar^{2+}$ and $Ar^{3+}$. However, in the integrated spectra of the knots only [ArIII] $\lambda$ 7136 Å was measured and the abundance of $Ar^{2+}$ was calculated assuming that T([ArIII]) $\approx$ T([SIII]) (Garnett 1992). As already described above, [ArIV] $\lambda$ 4741 Å was present in some knots and subtracted from their WR blue bump flux, but their contribution was small to yield any accurate abundance determination. The total abundance of Ar was then calculated using the ICF($Ar^{2+}$) derived from photo-ionization models by Pérez-Montero et al. (2007).

j Finally, for iron the [FeIII] $\lambda$ 4658 Å emission line was used together with the electron temperature of [OIII]. We have taken the ICF($Fe^{2+}$) from Rodríguez & Rubin (2004).

The ionic and total abundances –and their corresponding errors– for each observed element for knots A, B, and, C are given in Table 6.

### 6.3 Mass of the ionized gas

We have calculated the H$\alpha$ luminosities for the four knots from our observed values, correcting for extinction according to the values found from the spectroscopic analysis. The H$\alpha$ flux has been calculated in the mosaic, where no flux loss is expected. The resulting values are listed in Table 7.

The total number of hydrogen ionizing photons from the extinction corrected H$\alpha$ flux was computed using,

$$Q(H^0) = \frac{\alpha_B}{\alpha_{H\alpha}^{eff}} \times \frac{L_{H\alpha}}{h\nu_{H\alpha}}$$

From the number of Lyman continuum photons, Q($H^0$), the corresponding mass of ionized hydrogen, M(HII) is:

$$M_{HII} = Q(H^0) \frac{m_p}{n_e \alpha_B}$$

The values are also listed in Table 7. All of them, given the assumptions of no dust absorption, represent lower limits.

Table 7 also gives the estimated radius, in parsecs, of a circular aperture covering the spatial distribution of the fibers used in the integrated spectra of each knot. This value

**Table 3.** Observed and reddening corrected relative line intensities [F(Hβ)=I(Hβ)=1000] with their corresponding errors for the integrated spectrum of the PPak-field using the blue and red one-pointings (PPak-field Pointing-1), the blue mosaic (PPak-field Mosaic), and knots A, B, C and D using the blue and red one-pointing. The adopted reddening curve, $f(\lambda)$ (normalized to Hβ), the equivalent width of the emission lines, the reddening-corrected Hβ intensity ($I_{tot}$(Hβ) is the total intensity measured in the mosaic), and the reddening constant are also given.

| | | PPak-field Pointing-1 | | | PPak-field Mosaic | | |
|---|---|---|---|---|---|---|---|
| λ (Å) | f(λ) | F(λ) | EW(Å) | I(λ) | F(λ) | EW(Å) | I(λ) |
| 3727 [OII]$^a$ | 0.322 | 1995 ± 65 | 55.0 | 3459 ± 230 | 2078 ± 55 | 60.2 | 3568 ± 262 |
| 3868 [NeIII] | 0.291 | 146 ± 31 | 3.7 | 240 ± 53 | 128 ± 25 | 3.4 | 209 ± 43 |
| 4102 Hδ | 0.229 | 163 ± 16 | 4.9 | 241 ± 27 | 135 ± 18 | 4.0 | 198 ± 29 |
| 4340 Hγ | 0.157 | 426 ± 24 | 15.6 | 557 ± 42 | 420 ± 20 | 15.8 | 547 ± 42 |
| 4363 [OIII] | 0.149 | 2 ± 1 | 0.1 | 2 ± 1 | 2 ± 1 | 0.1 | 3 ± 1 |
| 4686 HeII | 0.050 | 9 ± 3 | 0.3 | 9 ± 4 | 8 ± 1 | 0.3 | 9 ± 1 |
| 4861 Hβ | 0.000 | 1000 ± 13 | 40.6 | 1000 ± 46 | 1000 ± 15 | 41.6 | 1000 ± 54 |
| 4959 [OIII] | -0.026 | 395 ± 12 | 16.0 | 378 ± 20 | 403 ± 12 | 16.6 | 386 ± 23 |
| 5007 [OIII] | -0.038 | 1198 ± 14 | 48.1 | 1123 ± 49 | 1214 ± 17 | 51.6 | 1139 ± 59 |
| 5876 HeI | -0.203 | 166 ± 11 | 7.8 | 117 ± 9 | 163 ± 20 | 7.8 | 116 ± 15 |
| 6548 [NII] | -0.296 | 309 ± 8 | 15.2 | 186 ± 8 | 333 ± 8 | 17.7 | 203 ± 9 |
| 6563 Hα | -0.298 | 4857 ± 9 | 238.6 | 2921 ± 90 | 4859 ± 7 | 257.6 | 2948 ± 107 |
| 6584 [NII] | -0.300 | 1123 ± 9 | 55.1 | 672 ± 21 | 1127 ± 12 | 59.8 | 681 ± 26 |
| 6678 HeI | -0.313 | 52 ± 7 | 2.6 | 31 ± 4 | 53 ± 7 | 2.7 | 31 ± 4 |
| 6717 [SII] | -0.318 | 690 ± 3 | 34.7 | 401 ± 12 | 683 ± 5 | 36.2 | 400 ± 14 |
| 6731 [SII] | -0.320 | 484 ± 3 | 24.4 | 280 ± 9 | 484 ± 6 | 25.7 | 283 ± 10 |
| 7065 HeI | -0.364 | 33 ± 7 | 1.5 | 18 ± 4 | ⋯ | ⋯ | ⋯ |
| 7136 [ArIII] | -0.374 | 112 ± 16 | 5.1 | 59 ± 9 | ⋯ | ⋯ | ⋯ |
| 7325 [OII]$^b$ | -0.398 | 76 ± 17 | 4.3 | 39 ± 9 | ⋯ | ⋯ | ⋯ |
| 9069 [SIII] | -0.594 | 460 ± 20 | 32.1 | 167 ± 8 | ⋯ | ⋯ | ⋯ |
| I(Hβ)(erg s$^{-1}$ cm$^{-2}$) | | 6.05 × 10$^{-13}$ | | | 1.72 × 10$^{-12}$ | | |
| c(Hβ) | | 0.74 ± 0.02 | | | 0.73 ± 0.02 | | |
| | | Knot A | | | Knot B | | |
| λ (Å) | f(λ) | F(λ) | EW(Å) | I(λ) | F(λ) | EW(Å) | I(λ) |
| 3727 [OII]$^a$ | 0.322 | 1730 ± 19 | 61.9 | 2883 ± 78 | 2782 ± 126 | 175.8 | 5085 ± 572 |
| 3868 [NeIII] | 0.291 | 97 ± 11 | 3.6 | 154 ± 18 | 121 ± 32 | 8.2 | 208 ± 58 |
| 4102 Hδ | 0.229 | 207 ± 4 | 9.2 | 298 ± 9 | 231 ± 38 | 22.7 | 354 ± 67 |
| 4340 Hγ | 0.157 | 395 ± 4 | 22.1 | 507 ± 12 | 500 ± 52 | 61.4 | 670 ± 92 |
| 4363 [OIII] | 0.149 | 5 ± 3 | 0.4 | 6 ± 4 | 2 ± 1 | 0.3 | 3 ± 2 |
| 4471 HeI | 0.115 | 33 ± 4 | 2.1 | 40 ± 5 | ⋯ | ⋯ | ⋯ |
| 4658 [FeIII] | 0.058 | 4 ± 1 | 0.3 | 5 ± 1 | ⋯ | ⋯ | ⋯ |
| 4686 HeII | 0.050 | 6 ± 2 | 0.4 | 6 ± 2 | ⋯ | ⋯ | ⋯ |
| 4861 Hβ | 0.000 | 1000 ± 5 | 77.9 | 1000 ± 19 | 1000 ± 23 | 99.3 | 1000 ± 81 |
| 4959 [OIII] | -0.026 | 538 ± 3 | 40.3 | 517 ± 10 | 329 ± 13 | 32.5 | 313 ± 27 |
| 5007 [OIII] | -0.038 | 1652 ± 6 | 123.4 | 1556 ± 28 | 938 ± 13 | 96.3 | 874 ± 67 |
| 5876 HeI | -0.203 | 160 ± 7 | 15.0 | 116 ± 5 | 157 ± 14 | 18.1 | 108 ± 12 |
| 6312 [SIII] | -0.264 | 8 ± 1 | 0.9 | 6 ± 1 | ⋯ | ⋯ | ⋯ |
| 6548 [NII] | -0.296 | 217 ± 4 | 23.5 | 136 ± 3 | 430 ± 13 | 60.8 | 247 ± 16 |
| 6563 Hα | -0.298 | 4634 ± 5 | 503.6 | 2890 ± 38 | 5090 ± 11 | 711.0 | 2915 ± 160 |
| 6584 [NII] | -0.300 | 801 ± 2 | 87.4 | 497 ± 7 | 1335 ± 10 | 184.0 | 760 ± 42 |
| 6678 HeI | -0.313 | 51 ± 3 | 5.6 | 31 ± 2 | 51 ± 12 | 6.6 | 29 ± 7 |
| 6717 [SII] | -0.318 | 380 ± 2 | 42.8 | 230 ± 3 | 722 ± 13 | 94.0 | 398 ± 22 |
| 6731 [SII] | -0.320 | 280 ± 2 | 31.6 | 169 ± 2 | 514 ± 13 | 67.4 | 282 ± 17 |
| 7065 HeI | -0.364 | 32 ± 3 | 3.3 | 18 ± 1 | 46 ± 15 | 6.6 | 23 ± 8 |
| 7136 [ArIII] | -0.374 | 131 ± 3 | 13.6 | 73 ± 2 | 109 ± 18 | 13.5 | 54 ± 9 |
| 7325 [OII]$^b$ | -0.398 | 54 ± 9 | 6.2 | 29 ± 5 | ⋯ | ⋯ | ⋯ |
| 9069 [SIII] | -0.594 | 561 ± 16 | 96.1 | 219 ± 6 | 437 ± 46 | 62.6 | 144 ± 16 |
| I(Hβ)(erg s$^{-1}$ cm$^{-2}$) | | 2.29 × 10$^{-13}$ | | | 1.37 × 10$^{-14}$ | | |
| I$_{tot}$(Hβ)(erg s$^{-1}$ cm$^{-2}$) | | 6.63 × 10$^{-13}$ | | | 3.69 × 10$^{-14}$ | | |
| c(Hβ) | | 0.69 ± 0.01 | | | 0.81 ± 0.03 | | |

$^a$ [OII] λλ 3726 + 3729; $^b$ [OII] λλ 7319 + 7330

is not intended to be accurate, but to provide an order of magnitude of the sizes involved.

As can be seen, the luminosities and ionized gas masses of H$^+$ from these four knots are typical of GEHRs (Kennicutt 1984; Diaz et al. 1991).

### 6.4 Properties of the WR population

In a coeval stellar cluster, Wolf-Rayet (WR) stars should be visible from about 2 Myr after the onset of star formation. They have a rather short life with the lowest mass ones dis-appearing at about ∼ 5 Myr. Therefore the existence of this WR stellar population can be very useful for the study and characterization of the ionizing stellar population in starburst galaxies (Mas-Hesse & Kunth 1991; Vacca & Conti 1992; Gonzalez-Delgado et al. 1994; Pérez-Montero & Díaz 2007). In addition to the initial mass of the star, another fundamental parameter in WR stars is its metallicity. Increasing the content of heavy elements, increases the mass loss rate due to winds (Maeder 1991; Crowther 2007). The more metal rich is the star, the lower will be the initial mass needed for it to become a WR star. The number of WR stars in a resolved cluster can be estimated just by counting them. For an unresolved cluster, however, this is not possible. In this case, the equivalent width of the blue bump (WR λ 4650) or the ratio between this bump and the Hβ emission line can be used as diagnostics. The values of these quantities in the range of interest (2-6 Myr) depend strongly on metallicity.

**Table 3.** (*Continued*)

|  |  | Knot C |  |  | Knot D |  |  |
|---|---|---|---|---|---|---|---|
| λ (Å) | f(λ) | F(λ) | EW(Å) | I(λ) | F(λ) | EW(Å) | I(λ) |
| 3727 [OII][a] | 0.322 | 2064 ± 51 | 93.2 | 3676 ± 215 | 1711 ± 86 | 54.2 | 3290 ± 299 |
| 3868 [NeIII] | 0.291 | 53 ± 16 | 1.8 | 89 ± 28 | 100 ± 29 | 4.5 | 180 ± 55 |
| 4102 Hδ | 0.229 | 181 ± 22 | 6.4 | 273 ± 36 | 189 ± 39 | 9.4 | 301 ± 66 |
| 4340 Hγ | 0.157 | 453 ± 20 | 20.7 | 600 ± 38 | 359 ± 23 | 20.1 | 494 ± 45 |
| 4363 [OIII] | 0.149 | 2 ± 1 | 0.1 | 2 ± 1 | ⋯ | ⋯ | ⋯ |
| 4658 [FeIII] | 0.058 | 5 ± 2 | 0.2 | 5 ± 2 | ⋯ | ⋯ | ⋯ |
| 4686 HeII | 0.050 | 4 ± 2 | 0.1 | 4 ± 2 | ⋯ | ⋯ | ⋯ |
| 4861 Hβ | 0.000 | 1000 ± 12 | 76.2 | 1000 ± 42 | 1000 ± 17 | 60.6 | 1000 ± 60 |
| 4959 [OIII] | -0.026 | 198 ± 6 | 8.6 | 189 ± 9 | 170 ± 13 | 11.2 | 161 ± 16 |
| 5007 [OIII] | -0.038 | 622 ± 9 | 27.5 | 581 ± 24 | 488 ± 14 | 31.5 | 452 ± 28 |
| 5876 HeI | -0.203 | 196 ± 16 | 8.7 | 136 ± 12 | 137 ± 31 | 9.5 | 91 ± 21 |
| 6548 [NII] | -0.296 | 428 ± 9 | 19.2 | 252 ± 9 | 469 ± 10 | 34.4 | 257 ± 12 |
| 6563 Hα | -0.298 | 4981 ± 8 | 223.9 | 2921 ± 82 | 5232 ± 11 | 382.2 | 2859 ± 115 |
| 6584 [NII] | -0.300 | 1378 ± 9 | 62.0 | 804 ± 23 | 1588 ± 9 | 115.8 | 863 ± 35 |
| 6678 HeI | -0.313 | 43 ± 4 | 2.0 | 25 ± 3 | 53 ± 11 | 3.7 | 28 ± 6 |
| 6717 [SII] | -0.318 | 727 ± 5 | 33.2 | 411 ± 12 | 808 ± 7 | 58.4 | 423 ± 17 |
| 6731 [SII] | -0.320 | 523 ± 5 | 23.9 | 295 ± 9 | 588 ± 7 | 42.6 | 307 ± 13 |
| 7065 HeI | -0.364 | 30 ± 6 | 1.2 | 16 ± 3 | 26 ± 6 | 1.7 | 12 ± 3 |
| 7136 [ArIII] | -0.374 | 82 ± 15 | 3.5 | 42 ± 8 | 76 ± 12 | 5.2 | 36 ± 6 |
| 7325 [OII][b] | -0.398 | 58 ± 13 | 2.7 | 29 ± 6 | ⋯ | ⋯ | ⋯ |
| 9069 [SIII] | -0.594 | 451 ± 26 | 24.0 | 156 ± 9 | 442 ± 22 | 89.1 | 132 ± 7 |
| I(Hβ)(erg s$^{-1}$ cm$^{-2}$) | | | 5.13 × 10$^{-14}$ | | | 1.88 × 10$^{-14}$ | |
| I$_{tot}$(Hβ)(erg s$^{-1}$ cm$^{-2}$) | | | 1.28 × 10$^{-13}$ | | | 5.28 × 10$^{-14}$ | |
| c(Hβ) | | | 0.78 ± 0.02 | | | 0.88 ± 0.02 | |

[a] [OII] λλ 3726 + 3729; [b] [OII] λλ 7319 + 7330

**Table 4.** Electron densities and temperatures for the integrated spectra of the PPak-field and the four knots. Densities in cm$^{-3}$ and temperatures in $10^4$ K.

|  | PPak-field | Knot A | Knot B | Knot C | Knot D |
|---|---|---|---|---|---|
| n([SII]) | 55: | 61 ± 20 | 98: | 77: | 73:[a] |
| t([OIII])[b] | 0.78 ± 0.06 | 0.87 ± 0.06 | 0.80 ± 0.13 | 0.78 ± 0.10 | ⋯ |
| t([OII]) | 0.83 ± 0.11 | 0.80 ± 0.07 | 0.87 ± 0.13[c] | 0.78 ± 0.15 | ⋯ |
| t([SIII]) | 0.61 ± 0.13[d] | 0.71 ± 0.12 | 0.63 ± 0.20[d] | 0.61 ± 0.17[d] | ⋯ |

[a] Assuming a mean temperature from the temperature measured in the other knots
[b] [OIII] temperatures measured in the mosaic
[c] From a relation with T([OIII]) based on photoionization models
[d] From an empirical relation with T([OIII])

We have detected and measured the WR blue bump in three of the four knots in our observed HII region, The red bump was marginally detected in knot A, indicating the existence of WC stars, but no measurements are listed due to its very low signal-to-noise.

We have compared the observed relative intensities and equivalent widths of the WR bumps with the predictions of Starburst99 (Leitherer et al. 1999) population synthesis models as a function of age and star formation mode of the cluster. We have run Starburst99 models using stellar model atmospheres from Smith et al. (2002), Geneva evolutionary tracks with high stellar mass loss (Meynet et al. 1994), a Kroupa Initial Mass Function (IMF; Kroupa 2002) in two intervals (0.1-0.5 and 0.5-100 M$_\odot$) with different exponents (1.3 and 2.3 respectively), wind model (Leitherer et al. 1992), a supernova cut-off of 8 M$_\odot$ and for a metallicity Z = 0.02 (Z$_\odot$), the value closest to the total oxygen abundances derived.

Figure 12 shows the predicted equivalent width and intensity, relative to Hβ, of the blue bump as a function of the cluster age for a metallicity Z = 0.02 ($Z_\odot$), the value closest to the total oxygen abundances derived. The (blue) solid lines represent in both plots the evolution of an instantaneous burst of star formation, while the (green) dashed lines correspond to continuous star formation history with a constant star formation rate. The (light blue) horizontal band in both panels represents the error band associated with the measurement of the appropriate quantity for knot A. We can see that in the instantaneous star formation scenario, the WR features appear during a time interval between 2 and 5 Myr and they reach higher intensities. On the other hand, for a continuous star formation history, the WR features appear at the same age and, despite reaching lower intensities, converge to a non-zero value at older ages (see green dashed line in Figure 12).

**Table 5.** Ionic chemical abundances for helium.

| | PPak-field | Knot A | Knot B | Knot C |
|---|---|---|---|---|
| $He^+/H^+$ ($\lambda 4471$) | ... | $0.079 \pm 0.010$ | ... | ... |
| $He^+/H^+$ ($\lambda 5876$) | $0.082 \pm 0.007$ | $0.083 \pm 0.004$ | $0.075 \pm 0.010$ | $0.094 \pm 0.009$ |
| $He^+/H^+$ ($\lambda 6678$) | $0.074 \pm 0.011$ | $0.078 \pm 0.005$ | $0.070 \pm 0.018$ | $0.060 \pm 0.007$ |
| $He^+/H^+$ ($\lambda 7065$) | $0.084 \pm 0.018$ | $0.079 \pm 0.007$ | $0.110 \pm 0.038$ | $0.074 \pm 0.016$ |
| $He^+/H^+$ (Adopted) | $0.080 \pm 0.007$ | $0.080 \pm 0.004$ | $0.076 \pm 0.035$ | $0.073 \pm 0.025$ |

**Table 6.** Ionic chemical abundances derived from forbidden emission lines, ICFs and total chemical abundances for elements heavier than helium.

| | PPak-field | Knot A | Knot B | Knot C |
|---|---|---|---|---|
| $12 + \log(O^+/H^+)$ | $8.50 \pm 0.26$ | $8.50 \pm 0.18$ | $8.56 \pm 0.24$ | $8.67 \pm 0.24$ |
| $12 + \log(O^{2+}/H^+)$ | $7.98 \pm 0.13$ | $7.89 \pm 0.10$ | $7.84 \pm 0.26$ | $7.69 \pm 0.21$ |
| **$12 + \log(O/H)$** | $8.62 \pm 0.23$ | $8.60 \pm 0.16$ | $8.63 \pm 0.24$ | $8.71 \pm 0.24$ |
| $12 + \log(S^+/H^+)$ | $6.40 \pm 0.15$ | $6.22 \pm 0.10$ | $6.34 \pm 0.14$ | $6.50 \pm 0.14$ |
| $12 + \log(S^{2+}/H^+)$ | $6.87 \pm 0.29$ | $6.79 \pm 0.17$ | $6.76 \pm 0.38$ | $6.84 \pm 0.35$ |
| $ICF(S^+ + S^{2+})$ | $1.01 \pm 0.11$ | $1.01 \pm 0.11$ | $1.01 \pm 0.12$ | $1.00 \pm 0.11$ |
| **$12 + \log(S/H)$** | $7.01 \pm 0.25$ | $6.89 \pm 0.16$ | $6.90 \pm 0.31$ | $7.01 \pm 0.29$ |
| $\log(S/O)$ | $-1.61 \pm 0.34$ | $-1.70 \pm 0.23$ | $-1.73 \pm 0.39$ | $-1.70 \pm 0.37$ |
| $12 + \log(N^+/H^+)$ | $7.34 \pm 0.15$ | $7.25 \pm 0.10$ | $7.34 \pm 0.14$ | $7.51 \pm 0.14$ |
| $ICF(N^+)$ | $1.30 \pm 1.03$ | $1.25 \pm 0.69$ | $1.18 \pm 1.00$ | $1.11 \pm 0.87$ |
| **$12 + \log(N/H)$** | $7.45 \pm 0.37$ | $7.35 \pm 0.26$ | $7.42 \pm 0.37$ | $7.55 \pm 0.37$ |
| $\log(N/O)$ | $-1.17 \pm 0.30$ | $-1.25 \pm 0.21$ | $-1.23 \pm 0.28$ | $-1.16 \pm 0.28$ |
| $12 + \log(Ne^{2+}/H^+)$ | $7.92 \pm 0.18$ | $7.49 \pm 0.13$ | $7.80 \pm 0.34$ | $7.49 \pm 0.29$ |
| $ICF(Ne^{2+})$ | $1.52 \pm 0.44$ | $1.65 \pm 0.38$ | $1.85 \pm 0.67$ | $2.54 \pm 1.30$ |
| **$12 + \log(Ne/H)$** | $8.10 \pm 0.22$ | $7.71 \pm 0.16$ | $8.08 \pm 0.39$ | $7.89 \pm 0.37$ |
| $\log(Ne/O)$ | $-0.52 \pm 0.24$ | $-0.89 \pm 0.18$ | $-0.57 \pm 0.43$ | $-0.82 \pm 0.39$ |
| $12 + \log(Ar^{2+}/H^+)$ | $6.37 \pm 0.31$ | $6.22 \pm 0.18$ | $6.27 \pm 0.41$ | $6.22 \pm 0.38$ |
| $ICF(Ar^{2+})$ | $1.22 \pm 0.07$ | $1.24 \pm 0.04$ | $1.23 \pm 0.09$ | $1.28 \pm 0.04$ |
| **$12 + \log(Ar/H)$** | $6.45 \pm 0.31$ | $6.31 \pm 0.18$ | $6.36 \pm 0.41$ | $6.33 \pm 0.38$ |
| $\log(Ar/O)$ | $-2.16 \pm 0.35$ | $-2.28 \pm 0.22$ | $-2.27 \pm 0.43$ | $-2.38 \pm 0.40$ |
| $12 + \log(Fe^{2+}/H^+)$ | ... | $5.64 \pm 0.14$ | ... | $5.94 \pm 0.27$ |
| $ICF(Fe^{2+})$ | ... | $1.41 \pm 0.14$ | ... | $1.35 \pm 0.13$ |
| **$12 + \log(Fe/H)$** | ... | $5.79 \pm 0.14$ | ... | $6.08 \pm 0.27$ |

The comparison of the WR relative intensities and equivalent widths of the three knots, shown in Table 2, indicates that all of them are larger than the model-predicted values for a constant star formation mode, but match up fairly well with the values predicted for an instantaneous burst of star formation. The only exception being knot B, where the WR equivalent width is slightly higher than predicted by models. However, the observed relative intensities are within the expected range of theoretical values. From both quantities, an age around 4 Myr provides a good fit for the three knots, indicating that these bursts are very young and almost coeval.

**Table 8.** Derived number of WR stars and ratios between WR and O stars.

| Object ID | N(WR) | WR/O |
|---|---|---|
| Knot A | $125 \pm 10$ | $0.047 \pm 0.008$ |
| Knot B | $5 \pm 2$ | $0.002 \pm 0.001$ |
| Knot C | $22 \pm 3$ | $0.008 \pm 0.001$ |

Table 8 shows the number of O and WR stars derived by comparing the dereddened luminosities of H$\alpha$ and of the WR blue bump with the predictions of the instantaneous model of Starburst99. It can be seen that knot A contains more than one hundred WR stars, as expected from the strong blue bump feature detected in its integrated spectra. Interestingly, Knot B has WR stars, despite showing only a weak contribution to the continuum map of Figure 4.

Another way to estimate the number of WR stars is by means of a calibration using data on individual WR stars. Smith (1991) estimates that a single WN7 star emits $3.2 \times 10^{36}$ ergs s$^{-1}$ at 4650Å, which would yield a total number of $120 \pm 10$ WR stars in knot A, very close to the value derived from the Starburst99 model. On the other hand, Vacca & Conti (1992) estimate that one WN7 star emits $1.7 \times 10^{36}$ ergs s$^{-1}$, thus rising the number of WR stars almost by a factor of 2.

**Table 7.** Hα flux and derived parameters for the (mosaic) integrated spectra of the four knots. All quantities have been corrected for reddening.

|        | Radius (pc) | F(Hα) (erg cm$^{-2}$s$^{-1}$) | L(Hα) (erg s$^{-1}$) | Q(H$^0$) (photon s$^{-1}$) | M(HII) (M$_\odot$) |
|--------|-------------|-------------------------------|----------------------|----------------------------|--------------------|
| Knot A | 190         | $1.93 \times 10^{-12}$        | $8.03 \times 10^{39}$| $7.9 \times 10^{51}$       | $2.5 \times 10^5$  |
| Knot B | 95          | $1.08 \times 10^{-13}$        | $4.52 \times 10^{38}$| $4.5 \times 10^{50}$       | $1.5 \times 10^4$  |
| Knot C | 115         | $3.76 \times 10^{-13}$        | $1.55 \times 10^{39}$| $1.5 \times 10^{51}$       | $5.0 \times 10^4$  |
| Knot D | 100         | $1.51 \times 10^{-13}$        | $6.28 \times 10^{38}$| $6.2 \times 10^{50}$       | $2.0 \times 10^4$  |

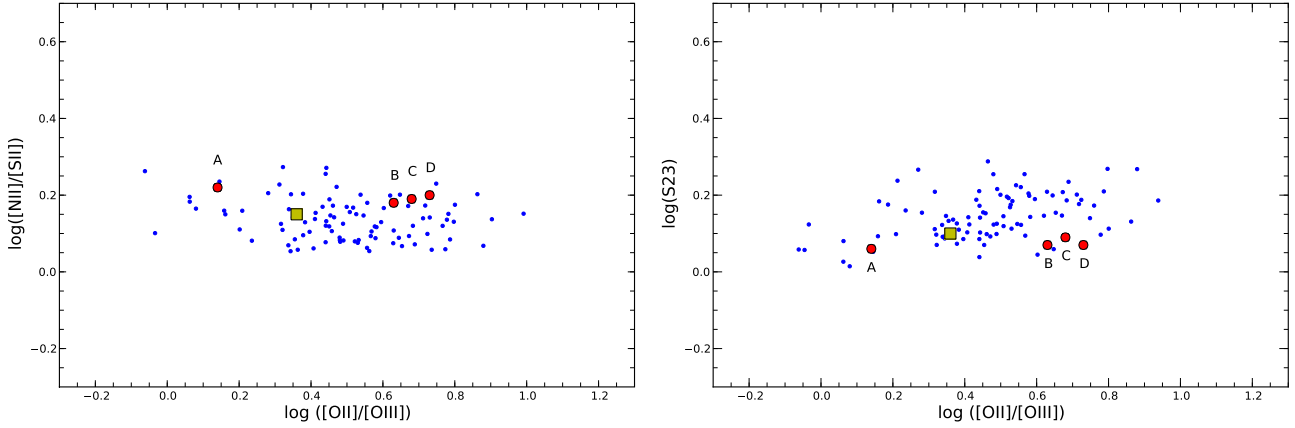

**Figure 13.** Point-to-point variations of the line ratio [NII]/[SII] (*left*) and S$_{23}$ parameter (*right*) versus the excitation measured as [OII]/[OIII]. The (red) circle points represent the integrated values of the knots, while the (yellow) square points the position in the diagram of the whole PPak field.

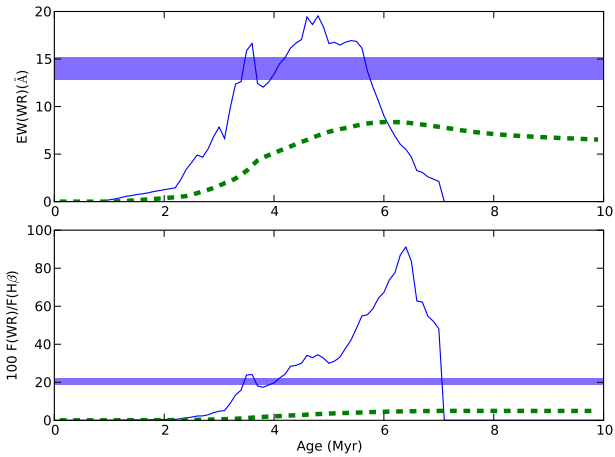

**Figure 12.** Relation between intensities and equivalent widths of the Wolf-Rayet blue bump as a function of cluster age for a Z = 0.02 metallicity, according to Starburst99 predictions. In both panels, (blue) solid lines represent instantaneous star formation and (green) dashed lines, continuous star formation. The (light blue) horizontal band represents the error in the measurement of the appropriate quantity for knot A.

## 7 DISCUSSION

### 7.1 Ionization structure

Line ratios that are relevant for the study of the ionization structure, as well as derived ionization parameters, are given in Table 9.

As discussed in Section 5.4, Figure 7 shows the BPT diagrams for the spectra. The (red) circle points represent the values of the knots, while the (yellow) square point is the position in the diagram of the whole PPak field. It can be seen that the integrated measurements span over the point-to-point values, with the same large spread in log([OIII] λ 5007/Hβ) and a relatively small variation in the other two line ratios. As expected, the integrated value of the PPak field appears as a relatively high excitation value, implying that Knot A weights on the integrated spectrum as the most important contribution.

The ionization structure of a nebula depends essentially on the shape of the ionizing continuum and on the nebular geometry. Figure 13 shows the point-to-point behaviour of the observed line ratios [NII] λλ 6548,6584/[SII] λλ 6717,6731 and the S$_{23}$ parameter as a function of the degree of excitation measured by [OII]/[OIII]. All values were obtained from the first pointing set, since the quantities represented involve the infrared sulphur lines. Although the first ratio contains only blue lines, we have decided to use also only the first dithering for consistency in the comparison. For each point, the accuracy of the ratios is typically better than 20 percent.

The ratio [NII] λλ 6548,6584/[SII] λλ 6717,6731 versus the excitation seems fairly constant, although a slight relative decrease with [OII]/[OIII] can be appreciated. This effect was also found for several HII regions in M101 by McCall et al. (1985) and in NGC 604 by Diaz et al. (1987). According to these authors, this can be understood as an excitation effect due to the lower ionization potential of S$^+$ with respect to N$^+$ combined with the relative low S$^+$/S$^{++}$ ratio.

The S$_{23}$ parameter (see map in Figure 14 and right panel of Figure 13) seems to remain uniform despite the large range of excitation values. Nevertheless, better signal-to-noise in the infrared sulphur lines would be desirable in order to definitely test this behaviour.

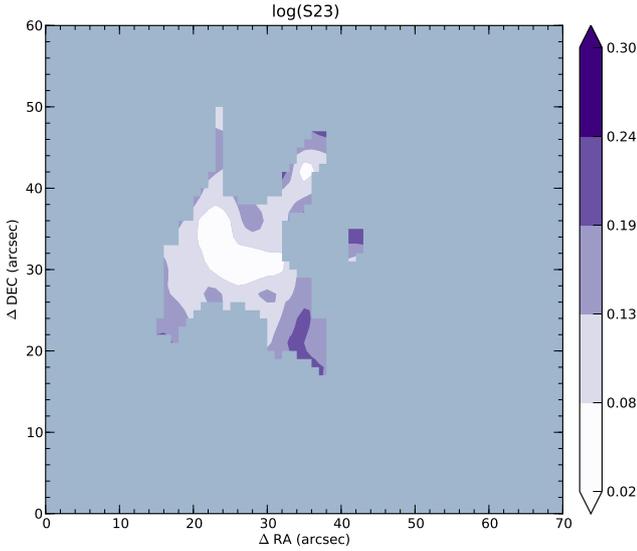

**Figure 14.** Map of the $S_{23}$ parameter.

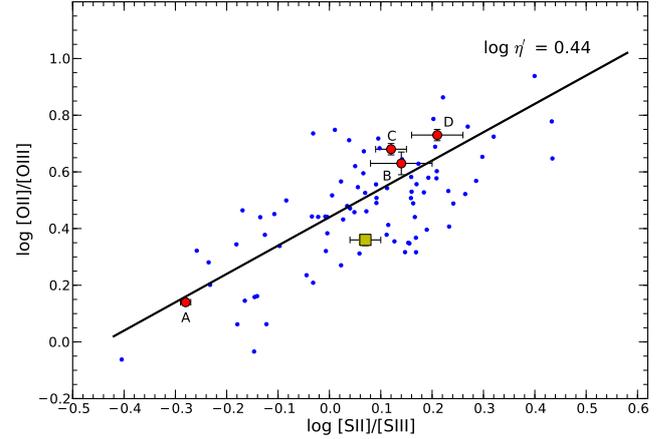

**Figure 15.** The $\eta$' plot: $log([O\textsc{ii}]/[O\textsc{iii}])$ vs. $log([S\textsc{ii}]/[S\textsc{iii}])$. Blue dots correspond to the different positions in the PPak field; red solid circles correspond to the four selected knots and the yellow square corresponds to the integrated PPak field. Diagonals in this diagram correspond to constant values of $\eta$' that are indicative of the temperature of the ionizing sources.

The volume-averaged ionization parameter $u$ is defined as

$$u = \frac{Q}{4\pi R^2 nc}$$

where $Q$ is the ionizing photon luminosity, $R$ is the radius of the Strömgren sphere, $n$ the number density of the gas, and $c$ the speed of light. This parameter is essentially the local ratio of Lyman continuum photons to gas density, which determines the degree of ionization at any particular location within the nebula. In principle, the ionization parameter can be estimated from the ratio of two lines of the same element in consecutive ionization states, such as [O\textsc{ii}]/[O\textsc{iii}] and [S\textsc{ii}]/[S\textsc{iii}]. Suitable expressions to be used can be deduced with the help of photoionization models.

$$log\, u = -1.68\, log([S\textsc{ii}]/[S\textsc{iii}]) - 2.99$$

for the sulphur lines Diaz et al. (1991), and

$$log\, u = -0.80\, log([O\textsc{ii}]/[O\textsc{iii}]) - 3.02$$

for the oxygen lines (Díaz et al. 2000).

For our studied region the values of both ratios found for each knot are typical of high metallicity H\textsc{ii} regions. Table 9 contains the ionization parameter as estimated from each of the ratios. As expected, the highest ionization parameter is found in Knot A.

As can be appreciated, the value of $u$ derived from the [O\textsc{ii}]/[O\textsc{iii}] ratio is systematically lower than the one obtained from the [S\textsc{ii}]/[S\textsc{iii}] ratio. However, the [OII]/[OIII] ratio depends on metallicity due to the presence of opacity edges of various abundant elements in the stellar atmospheres (see Díaz et al. 2007) with high metallicity regions showing higher [OII]/[OIII] ratios and hence apparently lower ionization parameters.

In fact, a plot of [OII]/[OIII] vs [SII]/[SIII] can provide important information about the temperature of the radiation field. The axes in this diagram define the $\eta$' parameter:

$$log\, \eta' \,=\, log\left[\frac{[O\textsc{ii}]\lambda\lambda\,3727,29\,/\,[O\textsc{iii}]\lambda\lambda\,4959,5007}{[S\textsc{ii}]\lambda\lambda\,6717,31\,/\,[S\textsc{iii}]\lambda\lambda\,9069,9532}\right]$$

This parameter is related to the softness parameter $\eta$ (Vilchez & Pagel 1988), through the electron temperature by the relation:

$$log\, \eta' \,=\, log\, \eta - \frac{0.14}{t_e} - 0.16$$

Figure 15 shows that the data corresponding to the different points in the PPak field cluster around a diagonal line corresponding to a constant value of log $\eta$' = 0.44, which indicates similar effective temperatures for the ionizing sources. This value is intermediate between the values shown by giant extragalactic HII regions in the discs of spirals and HII galaxies (Díaz et al. 2007). The data corresponding to the four knots (red solid circles) show a mean log $\eta$' value of 0.50 . However, the spectrum corresponding to the integrated PPak field (yellow square in the plot) would point to lower values of log $\eta$', 0.29, closer to those found in HII galaxies and indicating higher stellar effective temperatures. This might be revealing possible effects of lack of spatial resolution in extended GEHR that should be further explored.

These lack of resolution effects can also be seen in the hydrogen line equivalent widths. As mentioned above, the H$\beta$ luminosity of the emission knots amounts to about 50% of the total PPak field of view, with Knot A being the brightest. This knot has the highest excitation as characterised by the [O\textsc{ii}]/[O\textsc{iii}] ratio and dominates the excitation deduced for the PPak field integrated spectrum. On the other hand the H$\beta$ equivalent width for this entire region is about 40 Å, much lower than the ones measured for the individual knots that range from 60 to 100 Å.

### 7.2 Metal content

We derived oxygen, sulphur, nitrogen, neon, argon and iron chemical abundances from the integrated spectrum of the PPak-field and in three of the four knots, where the auroral lines of [OIII] (PPak-FOV, knots A, B, and C), [OII] (PPak-FOV, knots A and C) and [SIII] (knot A) were measured and thus their respective electron temperature derived. However, due to the quality of the data, the electron temperatures and metallicities obtained, have associated errors comparable to the dispersion found in empirical calibrations.

The temperatures (see Table 4) for the three knots and the integrated spectrum of the PPak-field show very similar values within the errors, around 8000K for T([OIII]). Similarly, all the values for T([OII]) are very close. The only exception is the one estimated for knot B from T([OIII]) based on photoionization models. This value is slightly higher than the mean value for the other knots, obtained from direct measurements. It is known that the relation between T([OII]) and T([OIII]) shows large scatter. The [OII] temperature derived from photoionization models by Pérez-Montero & Díaz (2003) covers a range of values for a sample of HII galaxies, GEHRs and diffuse HII regions in the Galaxy and the Magellanic Clouds, as shown in Hägele et al. (2006). Higher density models show lower values of T([OII]) for a given T([OIII]), being this effect more noticeable at high electron temperatures (see Figure 8 of Hägele et al. 2008). Our objects lie in the low temperature region, where $N_e = 100$ and $N_e = 500$ models are very close. T([OII]) derived from T([OIII]) is slightly higher than the derived directly, but still within the errors for our sample. The relation based on Stasińska (1990) photoionization models, could yield higher (by 2000 K) T([OII]) for the range in the models covered by our data. These differences translate into lower $O^+/H^+$ ratios, and therefore lower total abundances by an amount of 0.35 dex for knot A. The difference between the temperatures derived from Pérez-Montero & Díaz (2003) and those measured directly instead, give comparable abundances within the errors for the objects presented here.

Regarding the [SIII] temperature, the only direct value is that for Knot A (7100 ± 1200 K). The value derived from the empirical relation found by Hägele et al. (2006) using T([OIII]) is 7400 ± 1300 K, which matches the direct measurement. Following Garnett (1992) gives a higher [SIII] temperature (by 1400 K). As already mentioned we have used Hägele et al. (2006) to estimate T([SIII]) for the knots where the [SIII] λ 6312 Å auroral line could not be measured.

The abundances derived from the integrated spectra show values near solar. The abundances of various elements in HII regions in NGC 6946 have been studied by McCall et al. (1985) and Ferguson et al. (1998). At the galactocentric distance of the HII complex discussed in this paper, these studies also find close to solar oxygen abundances. Belley & Roy (1992) estimated the global oxygen abundance gradient of NGC 6946 by means of imaging spectrophotometry in the nebular lines Hα, Hβ, [NII], and [OIII]. Using the empirical calibration by Edmunds & Pagel (1984), they derived Δ log (O/H) / ΔR = -0.089 ± 0.003 dex kpc$^{-1}$, and an extrapolated central abundance of 12+log(O/H) = 9.36 ± 0.02. The same value was estimated by Ferguson et al. (1998) using emission line spectra and the empirical relation calibrated by McGaugh (1991).

To our knowledge, there is no other direct abundance determination in the literature for the HII complex presented here. Using our value and the value of the oxygen abundance gradient from Belley & Roy (1992), we find an extrapolated central abundance of 9.23 consistent with the values mentioned above. Another independent abundance determination for the disc of NGC 6946 has been reported by Larsen et al. (2006), obtained from H and K band spectra for a young luminous super star cluster at 4.8 kpc from the centre. This abundance, $12 + \log(O/H) = 8.76$, is about 0.2 dex lower than the value extrapolated from our data.

The logarithmic N/O ratios found for Knot A, B, and C are -1.25 ± 0.21, -1.23 ± 0.28 and -1.16 ± 0.28 respectively. They point to a constant value within the errors, close to the average value shown for disk HII regions at the same oxygen abundance (see Mollá et al. 2006). It is worth noting that, had we derived T([OII]) from t([OIII]) using Stasińska (1990) models, the N/O ratio would have been larger by a factor of about 2.

The log(S/O) values found for the three knots are -1.75 ± 0.23, -1.73 ± 0.39 and -1.70 ± 0.37, which are lower by a factor of 2.2 than log (S/O) = -1.39, the solar value (Grevesse & Sauval 1998), but consistent with the more recent solar values derived from 3D hydrodynamical models (log (S/O) = -1.60; Ludwig et al. 2009). Both the log(N/O) and log(S/O) ratios are very similar to the values found for the high metallicity HII regions analysed by Castellanos et al. (2002).

The abundances in knot D could only be derived through empirical calibrations based on strong emission lines since no auroral lines were detected. Different strong-line methods for deriving abundances, have been widely studied in the literature. Pérez-Montero & Díaz (2005) obtained different uncertainties for each oxygen abundance calibrator in a sample of ionized gaseous nebulae with accurate determinations of chemical abundances in a wide range of metallicity. We have applied these empirical calibrators to all the integrated spectra, whether or not the auroral lines have been detected. Among the strong-line parameters available, we have used the $O_{23}$ parameter (also known as $R_{23}$, originally defined by Pagel et al. (1979) and based on [OII] and [OIII] strong emission lines), which is characterised by its double-valued relation with metallicity. According to the values derived by the direct method in the other knots, we use the analytic expressions for the McGaugh (1991) upper branch given by Kobulnicky et al. (1999). We also used N2 defined by Storchi-Bergmann et al. (1994) and calibrated by Denicoló et al. (2002), based on the strong emission lines of [NII] which remains single-valued up to high metallicities; O3N2, defined by Alloin et al. (1979) and recently recalibrated by Pettini & Pagel (2004), which uses the strong emission lines of [OIII] and [NII]; $S_{23}$, defined by Vilchez & Esteban (1996) and calibrated by Díaz & Pérez-Montero (2000) and Pérez-Montero & Díaz (2005); $S_{23}/O_{23}$, defined by Díaz & Pérez-Montero (2000) and calibrated by Pérez-Montero & Díaz (2005) also showing a monotonical increase with oxygen abundance up to the oversolar regime; $S_3O_3$ and $Ar_3O_3$, defined and calibrated by Stasińska (2006). Finally the N2O2 parameter, defined by Pérez-Montero & Díaz (2005) as the ratio between [NII] and [OII] emission lines, can be used to obtain the N/O ratio. In this case, we have used the calibration by Díaz et al. (2007), which includes high metallicity HII regions.

The derived oxygen and nitrogen abundances and their uncertainties are presented in Table 9. The uncertainty is found from the standard deviation of the residuals of each parameter (see Pérez-Montero & Díaz 2005). As can be seen, the estimated oxygen abundances agree with those obtained from the direct method, all of them around the solar value. The $O_{23}$ ($R_{23}$) parameter provides the highest oxygen abundance, while the parameters involving the sulphur lines yield somewhat lower values.

**Table 9.** Oxygen and nitrogen abundances and their uncertainties for each integrated spectrum as derived using different empirical calibrators. The line ratios representative of the ionization structure, the ionization parameter ($u$) and the $\eta$ and $\eta'$ parameters, are also included.

|  | PPak-field | Knot A | Knot B | Knot C | Knot D |
|---|---|---|---|---|---|
| $\log(O_{23})$ | $0.70 \pm 0.03$ | $0.70 \pm 0.01$ | $0.80 \pm 0.05$ | $0.65 \pm 0.03$ | $0.59 \pm 0.04$ |
| $12 + \log(O/H)$ ($O_{23}$-upper) | $8.67 \pm 0.19$ | $8.69 \pm 0.19$ | $8.51 \pm 0.19$ | $8.69 \pm 0.19$ | $8.75 \pm 0.19$ |
| N2 | $-0.66 \pm 0.02$ | $-0.79 \pm 0.01$ | $-0.59 \pm 0.03$ | $-0.57 \pm 0.02$ | $-0.54 \pm 0.02$ |
| $12 + \log(O/H)$ (N2) | $8.49 \pm 0.25$ | $8.40 \pm 0.25$ | $8.55 \pm 0.25$ | $8.56 \pm 0.25$ | $8.60 \pm 0.25$ |
| O3N2 | $0.69 \pm 0.03$ | $0.96 \pm 0.01$ | $0.53 \pm 0.06$ | $0.32 \pm 0.03$ | $0.18 \pm 0.04$ |
| $12 + \log(O/H)$ (O3N2) | $8.51 \pm 0.25$ | $8.42 \pm 0.25$ | $8.56 \pm 0.25$ | $8.63 \pm 0.25$ | $8.67 \pm 0.25$ |
| $\log(S_{23})$ | $0.10 \pm 0.02$ | $0.06 \pm 0.01$ | $0.07 \pm 0.04$ | $0.09 \pm 0.02$ | $0.07 \pm 0.03$ |
| $12 + \log(O/H)$ ($S_{23}$) | $8.34 \pm 0.20$ | $8.27 \pm 0.20$ | $8.28 \pm 0.20$ | $8.33 \pm 0.20$ | $8.29 \pm 0.20$ |
| $\log(S_{23}/O_{23})$ | $-0.60 \pm 0.04$ | $-0.63 \pm 0.01$ | $-0.73 \pm 0.07$ | $-0.55 \pm 0.03$ | $-0.52 \pm 0.05$ |
| $12 + \log(O/H)$ ($S_{23}/O_{23}$) | $8.39 \pm 0.27$ | $8.34 \pm 0.27$ | $8.22 \pm 0.27$ | $8.45 \pm 0.27$ | $8.50 \pm 0.27$ |
| $\log(S_3O_3)$ | $-0.83 \pm 0.03$ | $-0.85 \pm 0.01$ | $-0.78 \pm 0.06$ | $-0.57 \pm 0.03$ | $-0.53 \pm 0.04$ |
| $12 + \log(O/H)$ ($S_3O_3$) | $8.43 \pm 0.25$ | $8.42 \pm 0.25$ | $8.45 \pm 0.25$ | $8.53 \pm 0.25$ | $8.54 \pm 0.25$ |
| $\log(Ar_3O_3)$ | $-1.28 \pm 0.07$ | $-1.33 \pm 0.01$ | $-1.21 \pm 0.08$ | $-1.14 \pm 0.08$ | $-1.10 \pm 0.07$ |
| $12 + \log(O/H)$ ($Ar_3O_3$) | $8.50 \pm 0.23$ | $8.46 \pm 0.23$ | $8.54 \pm 0.23$ | $8.58 \pm 0.23$ | $8.59 \pm 0.23$ |
| N2O2 | $-0.61 \pm 0.03$ | $-0.66 \pm 0.01$ | $-0.70 \pm 0.05$ | $-0.54 \pm 0.03$ | $-0.47 \pm 0.04$ |
| $\log(N/O)$ (N2O2) | $-1.18 \pm 0.14$ | $-1.22 \pm 0.14$ | $-1.25 \pm 0.14$ | $-1.14 \pm 0.14$ | $-1.09 \pm 0.14$ |
| $\log([OII]/[OIII])$ | $0.36 \pm 0.03$ | $0.14 \pm 0.01$ | $0.63 \pm 0.06$ | $0.68 \pm 0.03$ | $0.73 \pm 0.05$ |
| $\log(u)$ ([OII]/[OIII]) | $-3.31$ | $-3.13$ | $-3.53$ | $-3.56$ | $-3.60$ |
| $\log([SII]/[SIII])$ | $0.07 \pm 0.02$ | $-0.28 \pm 0.01$ | $0.14 \pm 0.04$ | $0.12 \pm 0.02$ | $0.21 \pm 0.02$ |
| $\log(u)$ ([SII]/[SIII]) | $-3.11$ | $-2.53$ | $-3.22$ | $-3.19$ | $-3.34$ |
| $\log([OIII]5007/H\beta)$ | $0.08 \pm 0.01$ | $0.22 \pm 0.01$ | $-0.03 \pm 0.01$ | $-0.21 \pm 0.01$ | $-0.31 \pm 0.01$ |
| $\log([SII]6717,31/H\alpha)$ | $-0.62 \pm 0.01$ | $-0.85 \pm 0.01$ | $-0.61 \pm 0.01$ | $-0.60 \pm 0.01$ | $-0.57 \pm 0.01$ |
| $\log([NII]/[SII])$ | $0.15 \pm 0.02$ | $0.22 \pm 0.01$ | $0.18 \pm 0.03$ | $0.19 \pm 0.02$ | $0.20 \pm 0.02$ |
| $\log(O^+/O^{2+})$ | $0.52 \pm 0.29$ | $0.61 \pm 0.20$ | $0.74 \pm 0.35$ | $0.97 \pm 0.32$ | $\cdots$ |
| $\log(S^+/S^{2+})$ | $-0.47 \pm 0.33$ | $-0.57 \pm 0.20$ | $-0.40 \pm 0.41$ | $-0.35 \pm 0.38$ | $\cdots$ |
| $\log(\eta)$ | $0.99 \pm 0.43$ | $1.18 \pm 0.29$ | $1.14 \pm 0.54$ | $1.32 \pm 0.50$ | $\cdots$ |
| $\log(\eta')$ | $0.29 \pm 0.04$ | $0.42 \pm 0.02$ | $0.49 \pm 0.07$ | $0.56 \pm 0.04$ | $0.52 \pm 0.05$ |

The N/O ratio obtained using the N2O2 parameter, gives similar values for the four knots, values which are also consistent with the ratio derived directly.

Both from the direct method and the empirical calibrators, we can conclude that there is no evidence for abundance variations in the four observed knots.

## 8 CONCLUSIONS

The recently developed technique of Integral Field Spectroscopy offers the opportunity to perform a spatially resolved study of the physical conditions of the ionized gas of GEHRs, exploring at the same time the properties of their ionizing stellar populations. This has been done in this work for the GEHR complex in NGC 6946 using PPak attached to the 3.5m telescope of the CAHA observatory. The configuration was chosen to cover the whole spectrum from 3600 up to 10000 Å, allowing the measurement of the [SIII] lines in the near infrared. This is the first time to our knowledge that data of this kind have been obtained for such a wide spectral range.

From the resulting maps, we have selected four main knots, labelled from A to D, to perform a detailed integrated spectroscopic analysis of these structures and of the whole PPak field-of-view. For all the knots the density has been found to be very similar and below 100 cm$^{-3}$. The [OIII] electron temperature was measured in knots A, B, C and the integrated PPak-field, and was found to be around 8000 K. The temperatures of [OII] and [SIII] were estimated in the four cases. The abundances derived using the "direct method" are typical of high metallicity disk HII regions, and very uniform among all knots, with a mean value of $12+\log(O/H)= 8.65$. This is comparable to what has been found in this galaxy by other authors for regions at similar galactocentric distance (Larsen et al. 2006). The S/O and N/O ratios are very similar in all structures. Therefore, a remarkable abundance uniformity is found despite the different excitations found throughout the nebula that can be appreciated in the [OIII] $\lambda$ 5007/H$\beta$, [NII] $\lambda$ 6584/H$\alpha$ and [SII] $\lambda\lambda$ 6717,6731/H$\alpha$ maps. This uniformity is also supported by the behaviour of the $S_{23}$ parameter which is approximately constant despite the observed wide range in excitation. This confirms the results found in the classical studies by Skillman (1985), Diaz et al. (1987), Rosa & Mathis (1987), and more recently in spatially resolved studies of HII galaxies (Kehrig et al. 2008; Lagos et al. 2009). On the other hand, there are stellar populations old enough as to produce supernovae since at least one is reported to exist in the region (Mayall 1948). This implies either a very fast and effective mixing with the surroundings or that, as suggested by Stasińska et al. (2007), the new metals processed and injected by the current star formation episode are not observed and reside in the hot gas phase. The metals from previous events, on the contrary, would be well mixed and homogeneously distributed through the whole extent of the region.

The Hβ luminosity of the knots amounts to about 50% of the total PPak field of view. Knot A is the brightest and provides 75% of the combined flux from the knots and 38% of the total Hβ flux. This knot has the highest excitation as characterised by the [OII]/[OIII] ratio and dominates the excitation deduced for the PPak field integrated spectrum. On the other hand the measured Hβ equivalent width for this entire region is about 40 Å, much lower than the ones measured for the individual knots that range from 60 to 100 Å. Some effects associated to the lack of spatial resolution could also be evidenced by the higher ionising temperature that would be deduced from the η' parameter measured in the integrated PPak spectrum with respect to the four studied emitting knots.

Wolf-Rayet features have been detected in knot A and C, and a weak contribution by these stars has also been found in knot B, leading to a derived total number of WR stars of 125, 22 and 5, for knots A, C and B respectively and provide O/WR number ratios which are consistent with SB99 models for an age of about 4 Myr for the metallicity of the region. Knot D, with no WR features, shows weak Hα emission, low excitation, and the lowest Hβ equivalent width, all of which points to a more evolved state.

## ACKNOWLEDGMENTS

We wish to express our gratitude to Sebastián Sánchez and Fabián Rosales for their invaluable help during the reduction process. This work has been partially supported by DG-ICYT grant AYA2007-67965-C03, AYA2007-67965-C03-02, AYA2007-64712, and Junta de Andalucía TIC 114. RGB acknowledges support from the Spanish MEC through FPI grant BES-2005-6910. ET, RT and JL thank the hospitality of the UAM during several visits when this work was developed. We thank an anonymous referee for very useful comments that improved the presentation of the paper.